\documentclass[prc,twocolumn,twoside,showpacs,floatfix]{revtex4}

\usepackage{graphicx,color,rotating}
\usepackage{amsmath,amssymb,bm}
\usepackage{dcolumn}

\usepackage{times,txfonts}

\newcommand{\eq}[1]{\begin{equation}#1\end{equation}}
\newcommand{\eqmulti}[1]{\begin{equation}\begin{split}#1\end{split}\end{equation}}

\newcommand{\ket}[1]{\ensuremath{\,|{#1}\rangle}}

\newcommand{\ketbra}[2]{\ensuremath{\,|{#1}\rangle\!\langle{#2}|\,}}
\newcommand{\matrixe}[3]{\ensuremath{\langle{#1}|\,{#2}\,|{#3}\rangle}}

\newcommand{\op}[1]{\ensuremath{#1}}

\renewcommand{\vec}[1]{\ensuremath{\mathbf{#1}}}

\newcommand{\vO}{\ensuremath{\op{v}}}

\newcommand{\HO}{\ensuremath{\op{H}}}

\newcommand{\TO}{\ensuremath{\op{T}}}

\newcommand{\VO}{\ensuremath{\op{V}}}
\newcommand{\WO}{\ensuremath{\op{W}}}

\newcommand{\MC}{\ensuremath{\mathcal{M}}}

\newcommand{\pOV}{\ensuremath{\vec{\op{p}}}}
\newcommand{\qOV}{\ensuremath{\vec{\op{q}}}}

\newcommand{\POV}{\ensuremath{\vec{\op{P}}}}

\newcommand{\XOV}{\ensuremath{\vec{\op{X}}}}

\newcommand{\UCOM}{\ensuremath{\textrm{UCOM}}}
\newcommand{\intr}{\ensuremath{\textrm{int}}}
\newcommand{\refe}{\ensuremath{\textrm{ref}}}
\newcommand{\cm}{\ensuremath{\textrm{cm}}}
\newcommand{\elem}[2]{\ensuremath{{}^{#2}\text{#1}}}

\newcommand{\fm}{\ensuremath{\,\text{fm}}}

\newcommand{\symboldiamond}[1][black]{{\color{#1}$\blacklozenge$}}
\newcommand{\symboltriangle}[1][black]{{\color{#1}$\blacktriangle$}}
\newcommand{\symbolbox}[1][black]{{\color{#1}$\blacksquare$}}
\newcommand{\symbolcircle}[1][black]{{\color{#1}$\bullet$}}

\newcommand{\symboldiamondopen}[1][black]{{\color{#1}$\lozenge$}}

\newcommand{\symbolboxopen}[1][black]{{\color{#1}$\square$}}
\newcommand{\symbolcircleopen}[1][black]{{\color{#1}$\circ$}}

\definecolor{FGViolet}{rgb}{0.61,0.32,0.61}
\definecolor{FGDarkBlue}{rgb}{0,0,0.6}
\definecolor{FGBlue}{rgb}{0,0,0.8}
\definecolor{FGLightBlue}{rgb}{0.2, 0.6, 0.8}
\definecolor{FGGreen}{rgb}{0.2,0.7,0.2}
\definecolor{FGLightGreen}{rgb}{0.4,1,0.4}
\definecolor{FGYellow}{rgb}{1,0.95,0}
\definecolor{FGOrange}{rgb}{0.95,0.5,0.1}
\definecolor{FGRed}{rgb}{0.8,0,0}
\definecolor{FGWhite}{rgb}{1,1,1}
\definecolor{FGLightGray}{rgb}{0.8,0.8,0.8}
\definecolor{FGGray}{rgb}{0.5,0.5,0.5}
\definecolor{FGDarkGray}{rgb}{0.3,0.3,0.3}
\definecolor{FGBlack}{rgb}{0,0,0}

\begin{document}

\title{Importance Truncation for Large-Scale Configuration Interaction Approaches}

\author{Robert Roth}
\email{robert.roth@physik.tu-darmstadt.de}

\affiliation{Institut f\"ur Kernphysik, Technische Universit\"at Darmstadt,
64289 Darmstadt, Germany}

\date{\today}

\begin{abstract}  

We introduce an iterative importance truncation scheme which aims at reducing the dimension of the model space of configuration interaction approaches by an \emph{a priori} selection of the physically most relevant basis states. Using an importance measure derived from multiconfigurational perturbation theory in combination with an importance threshold, we construct a model space optimized for the description of individual eigenstates of a given Hamiltonian. We discuss in detail various technical aspects and refinements of the importance truncation, such as perturbative corrections for excluded basis states, threshold extrapolation techniques, and different iterative model-space update schemes. We apply the idea of the importance truncation in the context of the  no-core shell model (NCSM) for the \emph{ab initio} description of nuclear ground states. In a series of benchmark calculations for closed- and open-shell nuclei up to \elem{O}{16} we compare the ground-state energies obtained in the importance truncated NCSM to the full NCSM. All calculations show an excellent agreement of importance truncated and full NCSM for all cases where the latter is feasible. The results demonstrate that the importance truncated NCSM, while preserving most of the advantages of the full NCSM, gives access to much larger $N_{\max}\hbar\Omega$ spaces and heavier nuclei. In this way we are able to perform importance truncated NCSM calculations for nuclei like \elem{C}{12} and \elem{O}{16} up to $N_{\max}=22$. 

\end{abstract}

\pacs{21.60.De, 21.60.Cs, 21.30.Fe, 02.70.-c}

\maketitle

\section{Introduction}

Configuration interaction (CI) approaches play an important role for the description of quantum many-body systems in many different areas of modern physics, ranging from atomic and molecular physics and quantum chemistry to condensed matter and nuclear physics. Well known examples for CI-type methods from the different fields include the full and truncated configuration interaction methods for the many-electron problem in molecular physics and quantum chemistry \cite{ShSc99}, the exact diagonalization approaches for Heisenberg- or Hubbard-type problems in condensed matter theory \cite{OuSc07,BePa00}, or the diagonalization shell model or general configuration-mixing approaches in nuclear structure physics \cite{CaMa05,NaQu09}. 

The basic framework of all of these methods is the same: Within a model space spanned by a set of many-body states, the eigenstates of the Hamiltonian are determined through a large-scale numerical solution of the matrix eigenvalue problem. The many-body states forming the basis of the model space are often Slater determinants of a set of single-particle states. The basic parameter which determines the difficulty and computational cost of such calculations is the dimension $D$ of the many-body model space, i.e., the linear size of the Hamilton matrix. If the full eigenspectrum is required, then exact numerical diagonalizations are routinely performed for dimensions up to $D\sim10^5$ nowadays. Often only a few eigenstates are of interest, such that Lanczos-type algorithms provide a very efficient tool and expand the domain of tractable model-space dimensions to $D\sim10^9$ \cite{CaMa05,NaQu09} and possibly $10^{10}$ through massive parallelization \cite{MaVa09}.

We consider applications which require only one or few low-lying eigenstates. In those cases the model space often contains a significant number of basis states that contribute to the basis expansion of the target eigenstates with extremely small or vanishing amplitudes. If these basis states would be omitted from the outset, the target eigenstates obtained by a solution of the eigenvalue problem in the truncated space and all observables derived from them would change only little. The diagonalization in the truncated space gives a variational approximation to the full eigenstates whose quality is directly controlled by the threshold on the amplitudes used to identify the important basis states. In order to exploit this idea, we need a way to estimate the amplitudes of the individual basis states without actually solving the full eigenvalue problem. This can be done in the framework of many-body perturbation theory, using the amplitudes for the first-order perturbative correction of an initial approximation for the target states as an importance measure \cite{RoNa07,RoGo08}. This is the concept of the importance truncation scheme discussed in this paper. 

We will focus on the nuclear many-body problem in the framework of a large-scale shell-model approach. However, all of the conceptual developments are generic and can be applied to any CI-type many-body method for other quantum systems as well.
For the nuclear many-body problem, we aim at an exact \emph{ab initio} solution for a Hamiltonian including a realistic nuclear interaction. In this context, the no-core shell model (NCSM) is the most successful CI-type method at present \cite{NaQu09,NaVa00,NaVa00b,NaKa00,NaOr02,NaOr03}. The model space of the NCSM is spanned by Slater determinants constructed from harmonic-oscillator single-particle states with an upper limit on the unperturbed excitation energy of the many-body basis states of $N_{\max}\hbar\Omega$. A unique advantage of the $N_{\max}\hbar\Omega$ truncation is the possibility to separate intrinsic and center-of-mass degrees of freedom and thus to obtain translationally invariant intrinsic states. The NCSM is able to provide a complete description of the ground and low-lying excited states including all relevant observables, like energies, transition matrix elements, form-factors and densities. It has been applied very successfully to nuclei up to mass $A\sim13$ using realistic Hamiltonians involving two- and three-nucleon interactions, e.g. the modern interactions derived within chiral effective field theory \cite{NaGu07,EnMa03,EpNo02}. A similar set of observables and nuclei is accessible in Green's function Monte Carlo calculations \cite{PiWi01,PiWi02,PiWi04} which, however, are restricted to certain classes of local interactions. Coupled-cluster methods, which have recently been used in connection with chiral two-nucleon interactions, have provided predictions also for heavier closed-shell nuclei \cite{HaPa08,HaDe07}. 

The range of applicability of the NCSM is limited solely by the combinatorial growth of the model space with particle number $A$ and energy truncation $N_{\max}\hbar\Omega$. For \elem{O}{16} the model space dimension reaches the order $10^{9}$ already for $N_{\max}=8$, which is typically not sufficient to obtain results that are converged with respect to $N_{\max}$. Since the model-space dimension is its only crucial limitation, the NCSM provides the optimal framework for implementing the importance truncation idea \cite{RoNa07}. As we will discuss in detail, the importance truncated no-core shell model (IT-NCSM) obtained in this way extends the NCSM to a much larger domain in $A$ and $N_{\max}$. 

This paper is organized as follows: In Sec. \ref{sec:its} we discuss the general elements of the importance truncation scheme that can be employed in any CI-type calculation. In Sec. \ref{sec:itncsm} we combine these elements with the NCSM and discuss the basic properties of the IT-NCSM. In Secs. \ref{sec:app} and \ref{sec:app2} we present a series of large-scale benchmark calculations in the IT-NCSM for ground states of different closed- and open-shell nuclei up to \elem{O}{16} and compare to the results of the full NCSM. Throughout this work we restrict ourselves to a regime where full NCSM calculation are still possible to some extent so that a detailed assessment of the importance truncation is possible.

\section{Importance Truncation Scheme}
\label{sec:its}

\subsection{Concept}
\label{sec:its_concept}

Consider a quantum many-body system whose ground and excited states shall be determined by solving the eigenvalue problem of the Hamiltonian in a large model space. The nuclear shell model is a typical example: The model space is spanned by a set of Slater determinants of harmonic oscillator single-particle states and the lowest few eigenvalues and the corresponding eigenvectors of the Hamilton matrix are determined. Similar configuration interaction (CI) methods are used throughout many fields of physics and chemistry. 
 
In all of these methods the many-body model space is constructed in a combinatorial fashion with some global truncation. In the no-core shell model in nuclear physics the model space is spanned by all possible Slater determinants constructed from harmonic oscillator single-particle states with total excitation energies up to $N_{\max}\hbar\Omega$. In a full configuration interaction calculation in quantum chemistry the model space is spanned by all Slater determinants that can be constructed from a given finite set of single-particle orbitals.

These global truncations do not account for the specific features of the Hamiltonian and the physical properties of the state one is interested in. As a result the model space contains a substantial number of basis states which are irrelevant for the description of a specific eigenstate, e.g., the ground state. The basic goal of the importance truncation scheme is to identify the important configurations for the description of one or a set of target states using the information provided by the Hamiltonian. Only the important states are selected to construct a new, greatly reduced model-space in which the eigenvalue problem is eventually solved. These importance selection ideas have been pioneered in quantum chemistry in the 1970s leading to a number of different computational schemes (see Sec. \ref{sec:its_othermethods}). The crucial ingredient is an \emph{a priori} measure for the importance of individual basis states. One possible framework to construct a simple importance measure is low-order multi-reference or multiconfigurational perturbation theory as discussed in the following section. Though the following is applicable to all types of configuration interaction approaches, we will employ the language of the nuclear shell model for convenience.

\subsection{Multiconfigurational Perturbation Theory}
\label{sec:its_perturbationtheory}

We start from a full model space $\MC_{\text{full}}$ spanned by a set of many-body basis states $\ket{\Phi_\nu}$---for example the harmonic-oscillator Slater determinants of the shell model with some model-space truncation. Furthermore we assume a reference state $\ket{\Psi_\refe}$ being a zeroth-order approximation for the eigenstate of the Hamiltonian we are interested in, e.g., the ground state. In general, the reference state can be a superposition of basis states from a subspace $\MC_\refe$ of the full model space
\eq{ \label{eq:its_referencestate}
  \ket{\Psi_\refe} = \sum_{\nu \in \MC_\refe} C_{\nu}^{(\refe)} \ket{\Phi_\nu} \;.
}
This initial approximation can be obtained, e.g., from a previous CI calculation for a smaller space. In the simplest case the reference space can be one-dimensional and the reference state is given by a single basis state $\ket{\Phi_0}$ corresponding, e.g., to the ground state of a closed shell nucleus in an independent-particle shell model.

Now we would like to use many-body perturbation theory to estimate the leading corrections to the reference state $\ket{\Psi_{\refe}}$ resulting from states outside of the reference space. Formally this requires the use of multireference or multiconfigurational perturbation theory (MCPT) as it is widely applied in quantum chemistry \cite{RoSz03,SuRo04}. 

For setting up the perturbation series we have to split the full Hamiltonian $\HO$ into an unperturbed part $\HO_0$ and a perturbation $\WO$. Since we want to start from the reference state $\ket{\Psi_{\refe}}$ as an unperturbed state, the unperturbed Hamiltonian has to be chosen such that 
\eq{
  \HO_0 \ket{\Psi_\refe} = \epsilon_\refe \ket{\Psi_\refe}
}
with an eigenvalue $\epsilon_\refe$ given by the expectation value with the full Hamiltonian $\HO$ 
\eq{
  \epsilon_\refe = \matrixe{\Psi_\refe}{\HO}{\Psi_\refe} \;.
}
Formally, we can write the unperturbed Hamiltonian which satisfies the eigenvalue relation as
\eq{ \label{eq:its_unperturbedhamiltonian}
  \HO_0 
  = \epsilon_\refe \ketbra{\Psi_\refe}{\Psi_\refe}
  + \sum_{\nu \notin \MC_\refe} \epsilon_\nu \ketbra{\Phi_\nu}{\Phi_\nu} \;.
}
For simplicity, contributions from states within $\MC_\refe$ which are orthogonal to $\ket{\Psi_\refe}$ have been left out, since they will not contribute later on. 
 
The unperturbed energies $\epsilon_\nu$ for basis states outside of the reference space $\MC_\refe$ which enter into the definition of the unperturbed Hamiltonian \eqref{eq:its_unperturbedhamiltonian} can be chosen freely. This choice of the unperturbed energies---and thus of the partitioning of the Hamiltonian---has an impact on the convergence behavior of the perturbation series and a number of different possibilities have been studied in this respect \cite{SuRo04}. In the simplest M\o{}ller-Plesset-type formulation of MCPT the unperturbed energies are defined as 
\eq{ \label{eq:its_moellerplesset}
  \epsilon_\nu = \epsilon_\refe + \Delta \epsilon_\nu \;,
}
where $\Delta \epsilon_\nu$ is the excitation energy of the basis state $\ket{\Phi_\nu}$ computed at the level of the independent-particle picture, i.e. using the single-particle energies of the underlying basis. When using a harmonic-oscillator basis, the single-particle energies are just the harmonic-oscillator energies $e_a = \hbar\Omega (2n_a + l_a + 3/2)$. When working with a Hartree-Fock single-particle basis, these are the Hartree-Fock single-particle energies.

Alternatively, in an Epstein-Nesbet partitioning, the unperturbed energies of states outside of the reference space are defined via the expectation value of the full Hamiltonian
\eq{
  \epsilon_\nu = \matrixe{\Phi_\nu}{\HO}{\Phi_\nu} \;,
}
which appears to be a more natural choice, but does not guarantee better convergence \cite{SuRo04}. 
For the present application computational efficiency is the prime concern, therefore the simple M\o{}ller-Plesset-type partitioning \eqref{eq:its_moellerplesset} is more appropriate and will be used eventually (cf. Sec. \ref{sec:itncsm_importancemeasure}). 

Once the unperturbed Hamiltonian is fixed, the perturbation $\WO$ is defined via
\eq{
  \WO = \HO - \HO_0
}
and we can easily write out the lowest orders of the Rayleigh-Schr\"odinger perturbation series.
For the energy the zeroth and first-order contributions read
\eqmulti{
  E^{(0)} 
  &= \matrixe{\Psi_{\refe}}{\HO_0}{\Psi_{\refe}} = \epsilon_\refe 
  \\
  E^{(1)} 
  &= \matrixe{\Psi_{\refe}}{\WO}{\Psi_{\refe}} = 0
}
as a direct consequence of our definition of the unperturbed Hamiltonian. The second-order contribution to the energy assumes the well known form
\eqmulti{ \label{eq:its_correctionenergy}
  E^{(2)} 
  &= -\sum_{\nu\notin\MC_\refe} \frac{ |\matrixe{\Phi_\nu}{\WO}{\Psi_\refe}|^2 }{\epsilon_\nu - \epsilon_\refe} \\
  &= -\sum_{\nu\notin\MC_\refe} \frac{ |\matrixe{\Phi_\nu}{\HO}{\Psi_\refe}|^2 }{\epsilon_\nu - \epsilon_\refe}  \;,
}
where we have used that all matrix elements of $\HO_0$ between $\ket{\Psi_\refe}$ and the basis states $\ket{\Phi_\nu}\notin\MC_\refe$ outside the reference space vanish by construction.
 
For the many-body states, the zeroth-order contribution is just given by the initial reference state
\eq{
  \ket{\Psi^{(0)}} = \ket{\Psi_\refe} \;.
}
The first-order correction is given by 
\eqmulti{ \label{eq:its_correctionstate}
  \ket{\Psi^{(1)}} 
  &= -\sum_{\nu\notin\MC_\refe} \frac{\matrixe{\Phi_\nu}{\WO}{\Psi_\refe}}{\epsilon_\nu - \epsilon_\refe}
    \ket{\Phi_\nu} \\
  &= -\sum_{\nu\notin\MC_\refe} \frac{\matrixe{\Phi_\nu}{\HO}{\Psi_\refe}}{\epsilon_\nu - \epsilon_\refe}
    \ket{\Phi_\nu} \;.
}
In all of these expressions we can insert the expansion \eqref{eq:its_referencestate} of the reference state $\ket{\Psi_\refe}$ in terms of the basis states. Obviously, all these relations reduce to ordinary many-body perturbation theory when dealing which a reference state that is given by a single basis state, i.e. for $\ket{\Psi_\refe} = \ket{\Phi_0}$.

In the following, MCPT serves two important purposes: (\emph{i}) It provides an efficient way to assess the importance of individual basis states outside of the reference space $\MC_\refe$ and will thus be the main ingredient in the importance truncation scheme. (\emph{ii}) It allows for a direct computation of corrections to the energy obtained by an initial shell-model calculation in a limited reference space $\MC_\refe$, induced by states outside of this simple space.

\subsection{Perturbative Importance Measure}
\label{sec:its_importancemeasure}
 
The central element of the importance truncation scheme is an \emph{a priori} measure for the relevance of individual basis states $\ket{\Phi_\nu}$ for the description of a specific eigenstate of the Hamiltonian. The target state is represented by an initial approximation, the reference state $\ket{\Psi_\refe}$, that carries the correct quantum numbers. Based on this reference state, multiconfigurational perturbation theory provides a natural framework for assessing the importance of basis states outside of the reference space $\MC_\refe$. 

A simple yet efficient importance measure can be constructed from the expression \eqref{eq:its_correctionstate} for the lowest-order correction to the unperturbed, i.e., reference state $\ket{\Psi_\refe}$. The amplitudes of the individual basis states $\ket{\Phi_\nu} \notin \MC_\refe$ in the perturbative correction \eqref{eq:its_correctionstate} provide a dimensionless measure for the relevance of those states. Thus we can use the perturbative amplitudes to define an \emph{a priori} importance measure:
\eqmulti{ \label{eq:its_importancemeasure}
  \kappa_{\nu} 
  &= -\frac{\matrixe{\Phi_\nu}{\HO}{\Psi_\refe}}{\epsilon_\nu - \epsilon_\refe} \\
  &= -\sum_{\mu\in\MC_\refe} C_{\mu}^{(\refe)} \frac{\matrixe{\Phi_\nu}{\HO}{\Phi_\mu}}{\epsilon_\nu - \epsilon_\refe} \;.
}
Only those basis states with an importance measure $|\kappa_\nu|$ larger than a threshold value $\kappa_{\min}$ are included in the importance-truncated model space. This space is tailored for an optimal description of the target state for the given Hamiltonian. In contrast to truncation schemes based on global energy cuts, the importance truncation criterion is directly governed by the Hamiltonian and the target state. The importance threshold $\kappa_{\min}$ controls the size of the model space and will later on be varied to investigate the dependence of the observables on the truncation. 

By construction, the importance measure $\kappa_{\nu}$ characterizes the basis states with regard to their relevance for the description of the eigenstate. This is not the only possible choice. One can define a corresponding importance measure for identifying the basis states which are most relevant for the description of the energy. Using the contributions of the individual basis states to the lowest-order correction to the energy \eqref{eq:its_correctionenergy} we can define the energy-based importance measure
\eqmulti{ \label{eq:its_importancemeasureenergy}
  \xi_{\nu} 
  &= -\frac{|\matrixe{\Phi_\nu}{\HO}{\Psi_\refe}|^2}{\epsilon_\nu - \epsilon_\refe} \;.
}
Since we are aiming at an optimum approximation to the eigenstate, which is then used for computing various observables other than the energy, the state-based importance measure $\kappa_{\nu}$ is conceptually superior and will be used in the following. In practice both measures lead to very similar results though the dimensionless state-based importance measure is easier to handle (cf. Sec. \ref{sec:itncsm_importancemeasure}).

It is important to note that for a two-body Hamiltonian the importance weight $\kappa_\nu$ (as well as $\xi_\nu$) vanishes whenever the basis state $\ket{\Phi_\nu}$ differs from all of the states in the reference space by more than two single-particle states. If we start from a single Slater-determinant as reference state, $\ket{\Psi_\refe}$, then only $1p1h$ and $2p2h$-excited states with respect to this determinant can yield non-zero matrix elements for $\HO$ and thus non-vanishing $\kappa_\nu$. In order to access $3p3h$ and $4p4h$-excited states directly, the second-order perturbative corrections to the amplitude would have to be used. This shows that the construction of the importance truncation via perturbation theory naturally entails a hierarchy of $npnh$ states. Only $1p1h$ and $2p2h$ excitations of $\ket{\Psi_\refe}$ contribute to the leading-order correction, $3p3h$ and $4p4h$ excited states first appear in the next-to-leading-order, and so on. In order to avoid the computationally demanding evaluation of higher-orders of perturbation theory we embed the first-order importance measure \eqref{eq:its_importancemeasure} into an iterative scheme for the construction of the importance truncated space as discussed in Sec. \ref{sec:its_modelspace}. 

Although we focus on the description of a single eigenstate, the concept of the importance truncation can easily be generalized to the simultaneous description of several eigenstates. Starting from a set of a few reference states $\ket{\Psi_{\refe}^{(n)}}$, we construct separate importance measures $\kappa_{\nu}^{(n)}$ for each reference state. The corresponding basis state $\ket{\Phi_\nu}$ is included into the importance truncated space if one of the importance measures $\kappa_{\nu}^{(n)}$ exceeds the threshold $\kappa_{\min}$, i.e. if the basis state contributes with a sizable amplitude to at least one of the target states. In this way, we obtain a model space tailored for the simultaneous description of all target states.

\subsection{Iterative Model-Space Construction}
\label{sec:its_modelspace}

Since the importance measure \eqref{eq:its_importancemeasure} constructed within lowest-order perturbation theory can only be used to extend the reference space by $1p1h$ and $2p2h$ excitations, we adopt an iterative procedure to construct the importance truncated model space for a given threshold $\kappa_{\min}$. Here we discuss a simple and universal update scheme applicable for any CI-type problem. More specialized update schemes can be devised for specific models spaces---we will come back to this question in the context of the NCSM in Sec. \ref{sec:itncsm_importanceupdate}.

Assume we start from a single basis state $\ket{\Phi_0}$ as an initial approximation for the target state, e.g., the ground state of a closed-shell nucleus. In the first iteration we use this state as reference state $\ket{\Psi_{\refe}^{[1]}} = \ket{\Phi_0}$ and employ the importance measure to construct all $1p1h$ and $2p2h$ excitations of the reference state with $|\kappa_\nu| \geq \kappa_{\min}$. Within this new model space $\MC^{[1]}(\kappa_{\min})$ consisting of up to $2p2h$ excitations we solve the eigenvalue problem and obtain an improved approximation for the target state 
\eq{
  \ket{\Psi^{[1]}} = \sum_{\nu\in\MC^{[1]}(\kappa_{\min})} C^{[1]}_{\nu} \ket{\Phi_\nu}
}
with amplitudes $C^{[1]}_\nu$ defined by the eigenvector. 

The improved state $\ket{\Psi^{[1]}}$ obtained in the first iteration is used to construct a new reference state $\ket{\Psi_{\refe}^{[2]}}$ for the second iteration. In order to accelerate the evaluation of the importance measure, we typically do not use the full eigenstate, but project onto a reference space $\MC^{[2]}_{\refe}$ spanned by the basis states $\ket{\Phi_{\nu}}\in\MC^{[1]}(\kappa_{\min})$ with amplitudes $C^{[1]}_{\nu}$ above a reference threshold, $|C^{[1]}_{\nu}|\geq C_{\min}$. The new reference state is thus defined as
\eq{
  \ket{\Psi_{\refe}^{[2]}} 
  = N_{\refe}^{[2]} \sum_{\nu\in\MC^{[2]}_{\refe}} C^{[1]}_{\nu} \ket{\Phi_\nu}
}
with a normalization constant $N_{\refe}^{[2]}$. Typically the reference threshold $C_{\min}$ can be chosen up to $10$-times larger than the importance threshold $\kappa_{\min}$ without affecting the results, we will discuss the threshold dependencies in detail later on. As in the first iteration, the importance measure is used to construct all $1p1h$ and $2p2h$ excitations with $|\kappa_\nu| \geq \kappa_{\min}$ on top of $\ket{\Psi_{\refe}^{[2]}}$. Since the new reference state already contains up to $2p2h$ excitations with respect to the initial Slater determinant, the model space $\MC^{[2]}(\kappa_{\min})$ consist of up to $4p4h$ excitations. From the solution of the eigenvalue problem we obtain a new approximation of the target state
\eq{
  \ket{\Psi^{[2]}} = \sum_{\nu\in\MC^{[2]}(\kappa_{\min})} C^{[2]}_{\nu} \ket{\Phi_\nu}
}
with new amplitudes $C^{[2]}_{\nu}$. This improved state again defines a new reference state and the previous steps are repeated. 

This scheme is used for a fully adaptive update of the whole model space, i.e., in each iteration the importance of all basis states is reassessed using the most recent reference state. In this way, the impact of the coupling to higher-order $npnh$ states is included when selecting states with lower $npnh$ orders. This relaxation can have sizable effects.

\subsection{\emph{A posteriori} corrections}
\label{sec:its_corrections}

Beyond the definition of the importance measure, perturbation theory can be used to construct \emph{a posteriori} corrections to the CI energies $E(\kappa_{\min})$, which account for contributions from basis states that are not included in the importance truncated model space $\MC(\kappa_{\min})$. We distinguish two types of corrections: (\emph{i}) those accounting for states that were discarded because of an importance measure below the threshold and (\emph{ii}) those accounting for configurations which would only be generated in the next iteration of the update cycle described in Sec. \ref{sec:its_modelspace} because of their $npnh$-order.

An estimate for the energy contribution of basis states with non-vanishing importance measure $|\kappa_{\nu}|<\kappa_{\min}$, i.e. those that were excluded from the importance truncated space $\MC(\kappa_{\min})$ for given threshold $\kappa_{\min}$, can be obtained from the second-order energy correction \eqref{eq:its_correctionenergy}. One can simply add the individual energy contributions of the basis states $\ket{\Phi_\nu}\notin\MC(\kappa_{\min})$:
\eq{ \label{eq:its_pertcorrectionexcluded}
  \Delta_{\text{excl}}(\kappa_{\min})
  = -\sum_{\nu\notin\MC(\kappa_{\min})}
    \frac{|\matrixe{\Phi_\nu}{\HO}{\Psi_\refe}|^2}{\epsilon_\nu - \epsilon_\refe} \;.
}
This amounts to adding the energy-importance measures $\xi_{\nu}$ defined in Eq. \eqref{eq:its_importancemeasureenergy} for the excluded configurations. Evaluating this correction during the construction of the importance truncated space does not generate any additional computational effort since the time-consuming matrix element has to be computed anyway for the importance measure $\kappa_{\nu}$. 

Generally, the correction $\Delta_{\text{excl}}(\kappa_{\min})$ provides only a rough estimate for the contribution of excluded states to the energy, since only the coupling to the reference state $\ket{\Psi_\refe}$ is considered but not the coupling to the majority of other basis states in $\MC(\kappa_{\min})$. The primary use of this correction relies on the formal property that $\Delta_{\text{excl}}(\kappa_{\min})$ has to vanish in the limit $\kappa_{\min}\to0$. This makes it a unique tool for stabilizing the extrapolation of the CI energy $E(\kappa_{\min})$ to vanishing importance threshold $\kappa_{\min}\to0$. This constrained threshold extrapolation is discussed in detail in Sec. \ref{sec:itncsm_thresholdextrapol}.

The effect of higher-order $npnh$ states that would only be generated in the next iteration of the model-space update can also be assessed via the second-order energy correction of MCPT. Assume we have performed two iterations of the importance-update cycle starting from a single Slater-determinant as initial reference state. The importance truncated space $\MC^{[2]}(\kappa_{\min})$ then contains up to $4p4h$ excitations with respect to the initial reference state. In order to estimate the effect of $5p5h$ and $6p6h$ configurations we can either perform a third iteration to construct $\MC^{[3]}(\kappa_{\min})$ and solve the CI problem or we can apply MCPT on top of the eigenstate $\ket{\Psi^{[2]}}$ obtained in the second iteration. Based on the second-order energy contribution given by Eq. \eqref{eq:its_correctionenergy}, we define the energy correction
\eqmulti{ \label{eq:its_pertcorrectionnextiteration}
  \Delta_{\text{PT}} 
  &= -\sum_{\nu\notin\MC_{\refe}} \frac{ |\matrixe{\Phi_\nu}{\HO}{\Psi_{\refe}}|^2 }{\epsilon_\nu - \epsilon_\refe}  \;,
}
where the reference state is given by the full eigenvector of the second iteration, $\ket{\Psi_{\refe}} = \ket{\Psi^{[2]}}$, and the sum runs over all $5p5h$ and $6p6h$ configurations. The computational effort for evaluating this correction is almost the same as a full iteration of the model-space update because of the complexity of the reference state. Reference thresholds and extrapolation techniques can be employed to speed up the calculations also in this case.

Simpler methods for estimating the effects of higher-order $npnh$ configurations are used in the context of truncated CI calculations in quantum chemistry \cite{ShSc99}. Due to their additional benefit of restoring size extensivity in truncated CI calculations they are commonly referred to as size-extensivity corrections \cite{LaDa74,DuDi94}. The simplest class of corrections are the single- or multi-reference Davidson corrections \cite{LaDa74}, which exist in a number of different formulations. In the language of quantum chemistry, a correction to the energy obtained in the second iteration corresponds to a multi-reference situation, where the eigenvector of the first iteration $\ket{\Psi^{[1]}}$ defines the reference state and the second iteration includes singles and doubles excitations on top of this reference state. Out of the different forms of multi-reference Davidson (MRD) corrections we use the so-called Davidson-Silver or Siegbahn form \cite{DaSi77,Sieg78,DuDi94}, which can be constructed in the context of perturbation theory,
\eqmulti{ \label{eq:its_mrdavidsoncorrection}
  \Delta_{\text{MRD}} 
  &= \Delta E_{21} \frac{1-C_{21}^2}{2 C_{21}^2 -1} \;,
}
where $E_{21} = E^{[2]} - E^{[1]}$ is the difference of the CI energies obtained in the second and the first iteration and 
\eq{
  C_{21}^2
  = \sum_{\nu \in \MC^{[1]}} |C^{[2]}_{\nu}|^2 
}
is the total weight with which the configurations in $\MC^{[1]}$, i.e. those that were already present in the first iteration, contribute to the eigenstate after the second iteration. Obviously the evaluation of the MRD correction does not involve any additional computational effort. For each value of the importance threshold $\kappa_{\min}$ we can extract the correction $\Delta_{\text{MRD}}(\kappa_{\min})$ using the energies and amplitudes of the two last iterations. Eventually the MRD correction is also extrapolated to vanishing threshold $\kappa_{\min}\to0$.

\subsection{Properties of the Importance Truncated CI}
\label{sec:its_properties}

Already at this stage we can identify a few general properties of the importance truncated CI, which do not depend on the details of the physical system or the model space under consideration.

First of all, it is a strictly variational approach. Since we determine energies always from a solution of an eigenvalue problem of the Hamiltonian in a restricted space, the lowest eigenvalue always provides an upper bound for the exact ground state energy. Moreover, the  Hylleraas-Undheim theorem \cite{HyUn30} applies, i.e., the energy of all states is guaranteed to drop monotonically with decreasing $\kappa_{\min}$ and is bounded from below by the exact eigenvalue $E_{n}^{\text{exact}}$ in the full model space: 
\eq{
  E_{n}^{\text{exact}} \leq E_{n}(\kappa_{\min}) \leq  E_{n}(\kappa'_{\min})
  \quad\text{for}\quad \kappa_{\min}<\kappa'_{\min} \;,
}
where $E_{n}(\kappa_{\min})$ is the $n$th energy eigenvalue obtained in the importance truncated space $\MC(\kappa_{\min})$. One can view the whole importance-truncated CI scheme as a variational calculation with an iteratively improved linear trial state. The set of states from which the trial state is constructed as a linear superposition, is selected using the importance measure based on a previous approximation of the target state.

Second, the iterative construction of the importance truncated model space will recover the full model space in the limit $(\kappa_{\min},C_{\min})\to0$ after $n/2$ iterations, where $n\leq A$ is the maximum $npnh$ excitation possible in the full model space, when starting with a single basis determinant as initial reference. As we will discuss in the context of the NCSM, more elaborate choices of the reference state will guarantee that this holds even after a single iteration. Together with the monotonous behavior of the energy, this limiting property provides the foundation for an \emph{a posteriori} extrapolation of the energies for different importance thresholds towards $\kappa_{\min}\to0$.

Third, the importance measure \eqref{eq:its_importancemeasure} is constructed to identify states based on their contribution to the expansion of the eigenstates and not based on their effect on the energies. Thus the importance truncation using $\kappa_{\nu}$ is tailored to generate an optimal approximation for the eigenstates in a limited model space. The energy can be computed from the eigenstates just like any other observable of interest. Therefore, from the conceptual point of view, all observables are accessible with the same precision as the energy.  

Finally, an interesting and nontrivial question that was raised in Refs.~\cite{DeHa08,RoNa08} and addressed in detail in Ref.~\cite{RoGo08} concerns the size extensivity of importance-truncated CI calculations. In simple terms, size extensivity requires that the energy obtained in a many-body calculation for a system composed of two non-interacting subsystems is equal to the sum of the energies obtained in separate calculations for the individual subsystems. Whereas full CI is size extensive, a truncation of the space at some fixed $npnh$ excitation level destroys size extensivity \cite{DuDi94,RoGo08}. Therefore, importance truncated CI calculations based on very few iterations of the model-space update discussed in Sec. \ref{sec:its_modelspace} can violate size extensivity. As discussed in detail in Ref. \cite{RoGo08} a computationally simple way to restore approximate size extensivity are Davidson-type corrections as given by Eq. \eqref{eq:its_mrdavidsoncorrection}. In most cases the effect of these corrections is small already after two iterative updates of the importance truncated space (cf. Sec. \ref{sec:app_O16}). After a sufficiently large number of iterations, i.e. once the model-space updates have converged, these size-extensivity corrections \eqref{eq:its_mrdavidsoncorrection} vanish altogether. This is in line with the fact, that after $A/2$ iterations at most the importance truncated CI recovers the full model space in the limit $(\kappa_{\min},C_{\min})\to0$ and thus would be manifestly size extensive. Although the limit of vanishing thresholds is realized only through an extrapolation, we can nevertheless presume that the importance truncated CI provides an approximately size-extensive result after convergence of the model-space updates and threshold extrapolation, simply because it provides an approximation of full CI without any explicit $npnh$ truncation.

\subsection{Comparison with other methods}
\label{sec:its_othermethods}

The idea of an importance selection was pioneered in quantum chemistry. Already in the late 1960s and early 1970s perturbative importance measures and thresholds were used to facilitate large-scale CI calculations \cite{WhHa69,HaWh71}. In a set of seminal papers Buenker and Peyerimhoff \cite{BuPe74,BuPe75,BuPe78} introduced a configuration-selecting multi-reference double-excitation CI approach (MRD-CI), which is  one of the benchmark methods in quantum chemistry up to today. It starts from a multi-configurational reference space and adds individual singles and doubles excitations employing a selection criterion based on the energy-lowering capability of the new configuration. The latter can be quantified either by using the perturbative second-order energy contribution \eqref{eq:its_importancemeasureenergy} or by explicitly evaluating the change of the energy eigenvalue obtained from adding the respective configuration. A threshold value on this energy-lowering is used to select the important configurations which are then included in the model space. Already in the initial applications of this MRD-CI scheme in Refs. \cite{BuPe74,BuPe75,BuPe78}, powerful threshold extrapolation techniques were employed to correct for the effects of excluded configurations (cf. Sec.~\ref{sec:itncsm_thresholdextrapol}). Moreover, size-extensivity corrections as discussed in Sec. \ref{sec:its_corrections} can be considered. 

Essentially all conceptual elements of the IT-CI scheme are already present in the MRD-CI (although we learned of the MRD-CI only after \cite{RoNa07} was published). One difference, however, lies in the iterative setup we adopt for the IT-CI which allows for a systematic improvement of the importance-truncated space. Whereas the MRD-CI is typically implemented as a one-step calculation, the idea of an iterative improvement of the model space has also been used in quantum chemistry. An example is the CIPSI method \cite{HuMa73,EvDa83,CiPe87} which uses a CI calculation for a limited model space of important configurations and supplements it with a second-order perturbative correction for singles and doubles excitations on top of the CI model space. The CI space is then iteratively enlarged by including those singles and doubles which contribute to the first-order perturbed states with amplitudes larger than a threshold value. Also this CIPSI scheme contains many of the relevant ideas employed in the IT-CI. 

Since these early formulations a large number of new implementations and variations of the aforementioned importance-selection ideas have been developed \cite{ShSc99,Harr91,AnCi97} and are being used for the \emph{ab initio} description of highly correlated problems in quantum chemistry. 

In nuclear physics the use of importance-selection techniques is not as far developed as in quantum chemistry. However, there are some schemes, particularly in the context of the valence-space shell model, which employ similar ideas. Among those is the Monte-Carlo Shell Model (MCSM) of Otsuka \emph{et al.} \cite{OtHo01}. It uses the lowering of the energy eigenvalue caused by adding a test configuration to a set of reference states as a criterion for the relevance of this configuration. However, the crucial element of this method is that the test configurations are generated through an imaginary time-evolution of the reference set implemented via an auxiliary-field Monte Carlo scheme. Due to this stochastic sampling the individual configurations are no simple shell-model basis  states anymore, but more complex states containing information on the Hamiltonian already. For the final diagonalization, typically supplemented by an angular momentum projection, a small number of those MCSM configurations is sufficient to capture the relevant physics. 

Another importance sampling scheme has been proposed by Andreozzi \emph{et al.} \cite{AnLo03} in connection with an iterative method for the solution of the eigenvalue problem \cite{AnPo02}. Here the approximations of the eigenvalues obtained during the iterative solution are used to apply an energy threshold criterion to discard irrelevant states. Horoi \emph{et al.} have devised a truncation scheme based on the diagonal matrix elements of the Hamiltonian and applied it in sd and fp-shell calculations \cite{HoBr94}.

\section{Importance Truncated No-Core Shell Model}
\label{sec:itncsm}

As the primary application we study the importance truncation scheme in connection with the no-core shell model (NCSM) \cite{RoNa07}. Applications of the importance truncation in nuclear CI approaches based on a different definition of the full model space have been presented in Ref. \cite{RoGo08}.

\subsection{Model space}
\label{sec:itncsm_modelspace}

The NCSM is based on an expansion of the many-nucleon state in a basis of Slater determinants of harmonic oscillator single-particle states. The model space of the full NCSM is restricted solely with regard to the maximum number of harmonic-oscillator excitation quanta, $N_{\max}$, in the many-body basis state. In other words, all harmonic-oscillator Slater determinants with unperturbed excitation energies of up to $N_{\max}\hbar\Omega$ are included in the model space. 

The combination of harmonic oscillator basis and $N_{\max}\hbar\Omega$ truncation has a unique advantage. Only this model space allows for an exact separation of the center-of-mass and intrinsic component of the many-body state for all $N_{\max}$. Therefore, one can guarantee that the intrinsic part of the state is free of spurious center-of-mass contaminations. Any other single-particle basis, e.g. a Hartree-Fock basis, or a different model-space truncation, e.g. a truncation at the level of the single-particle states like in other CI methods, will destroy this property and induce center-of-mass contaminations of the eigenstates which can severely affect intrinsic observables.  

The dimension of the $N_{\max}\hbar\Omega$ model space grows factorially with $N_{\max}$ and particle number $A$. Therefore, full NCSM calculations are computationally feasible only for relatively light nuclei or in very small spaces. Model space dimensions of the order of $10^9$ are used routinely with present NCSM codes \cite{CaNo99,MaVa09}. For \elem{O}{16} this allows for calculations in an $8\hbar\Omega$ space, which for most realistic Hamiltonians is not sufficient to reach convergence. The dimension of the $10\hbar\Omega$ model space is larger than $10^{10}$ and thus just beyond the reach of the full NCSM at present. For heavier nuclei the situation becomes progressively worse.

The importance truncation can be used to efficiently reduce the dimension of the $N_{\max}\hbar\Omega$ model space to a tractable size. Note that the $N_{\max}\hbar\Omega$ space already reflects a simplistic importance selection of the individual many-body basis states. Based on the perturbative arguments of Sec. \ref{sec:its_importancemeasure}, the amplitudes of basis states with large unperturbed excitation energies will be suppressed by the energy denominator in \eqref{eq:its_importancemeasure}. Precisely those states are discarded through the $N_{\max}\hbar\Omega$ truncation. However, the numerator of \eqref{eq:its_importancemeasure} and thus the full Hamiltonian, is not considered in this simplified picture. The $N_{\max}\hbar\Omega$ model space is not adapted to the specific properties of the Hamiltonian or the target states under consideration. By using the importance truncation in combination with the $N_{\max}\hbar\Omega$ model space we also include these aspects.

\subsection{Hamiltonian}
\label{sec:itncsm_hamiltonian}

For the following discussion we use a translationally invariant Hamiltonian composed of intrinsic kinetic energy $\TO_{\intr}=\TO-\TO_{\cm}$ and a realistic two-nucleon interaction $\VO_{\text{NN}}$:
\eq{ \label{eq:itncsm_hamiltonian}
  \HO_{\intr} 
  = \TO_{\intr} + \VO_{\text{NN}}
  = \frac{2}{A} \frac{1}{2\mu} \sum_{i<j}^{A} \qOV^2_{ij} + \sum_{i<j}^{A} \vO_{ij} \;,
}  
where $\qOV_{ij}=\frac{1}{2}(\pOV_i -\pOV_j)$ is the relative two-body momentum operator and $\mu=m_{\text{N}}/2$ the reduced mass. 

In principle any two-body interaction can be used as input. In this work we restrict ourselves to unitarily transformed interactions derived in the framework of the Unitary Correlation Operator Method (UCOM). Starting from the Argonne V18 potential a unitary transformation is used to account for short-range central and tensor correlations leading to a phase-shift equivalent effective interaction with improved convergence properties. The conceptual details of the UCOM approach are discussed in Refs. \cite{RoNe04,NeFe03,FeNe98}. Further details regarding the calculation of matrix elements of the $\VO_{\UCOM}$ interaction and the determinantion of the optimal correlation functions are discussed in Ref. \cite{RoHe05}. 

For all of the following calculations we use the `standard' set of correlation functions introduced in Ref. \cite{RoHe05} with a triplet-even tensor correlator with range parameter $I_{\vartheta}=0.09\fm^3$. This value was chosen such that experimental binding energies for \elem{H}{3} and \elem{He}{4} are roughly reproduced in full NCSM calculations. Though improved correlation functions are available \cite{RoRe08}, there exists a number of different many-body calculations for this first-generation $\VO_{\UCOM}$ interaction. Calculations for light nuclei in the NCSM and other methods \cite{Bacc07} show that $\VO_{\UCOM}$ exhibits good convergence properties and provides a realistic description of a number of observables. Studies of heavier nuclei in Hartree-Fock plus second-order many-body perturbation theory demonstrate that this interaction provides reasonable binding energies throughout the whole nuclear mass range without the explicit inclusion of a three-body interaction \cite{RoPa06}. Therefore the $\VO_{\UCOM}$ interaction provides a realistic testbed for the many-body methods investigated here. 

We emphasize that all of the following calculations use the Hamiltonian \eqref{eq:itncsm_hamiltonian} without further transformations, i.e., there is no additional Lee-Suzuki similarity transformation as in the \emph{ab initio} NCSM \cite{NaGu07,FoVa08,NaOr02,NaOr03,CaNa02,NaVa00,NaVa00b}. Here the term NCSM solely refers to a CI-type calculation specifically using an $N_{\max}\hbar\Omega$ model space.

\subsection{Implementation}
\label{sec:itncsm_implementation}

The implementation of an importance truncated NCSM differs from a conventional NCSM code. The computationally most demanding part is the construction of the importance truncated space itself. Due to the reduction of the dimension of the model space, the subsequent computation of the Hamilton matrix and the solution of the eigenvalue problem are simpler than in a full NCSM approach.

For generating the importance truncated space for a given reference state $\ket{\Psi_{\refe}}$, we use an algorithm motivated by the structure of the importance measure \eqref{eq:its_importancemeasure}. We loop over all basis states $\ket{\Phi_{\mu}}\in\MC_{\refe}$ contained in the reference state $\ket{\Psi_{\refe}}$ and create all $1p1h$ and $2p2h$ excitations of each of them. In order to avoid creating duplicates, we discard any newly created determinant $\ket{\Phi_{\nu}}$ that has a non-vanishing matrix element of the Hamiltonian with any of the states in $\MC_{\refe}$ that were processed previously. This update scheme, which is also used in importance selecting CI approaches in quantum chemistry \cite{HuMa73,BuPe78,Harr91}, is much more efficient than the simple scheme employed in \cite{RoNa07}. There, explicit loops over all possible $npnh$ excitations of the independent-particle shell-model state $\ket{\Phi_0}$ were used to generate candidate states for evaluating the importance measure. Though duplicates are excluded from the outset, this scheme becomes less efficient already at the $3p3h$ order and it eventually limited the calculations in Ref. \cite{RoNa07} to states up to the $4p4h$ level. Therefore, all results presented here are based on the refined implementation without any explicit limitation of the $npnh$ level of the states considered.

Evidently, the cost for the model-space update grows quadratically with the number of basis states in the reference $\ket{\Psi_{\refe}}$. Therefore, as discussed in Sec. \ref{sec:its_modelspace}, we introduce an additional reference threshold and define the reference state $\ket{\Psi_{\refe}}$ using the dominant components of the previous eigenstate. Typical reference thresholds $C_{\min}$ are of the order of $10^{-4}$ which leads to reference states composed of typically $10^5$ basis determinants. We always check that a further lowering of the reference threshold does not produce sizable effects. 

Eventually we obtain a list of basis states spanning the importance truncated model space including their importance weights. The typical dimensions we deal with are of the order $10^7$. These problems can be handled by conventional Lanczos- or Arnoldi-type algorithms---in addition to simple Lanczos-implementations we use the implicitly restarted Arnoldi algorithm of the ARPACK library \cite{ARPACK}. The many-body matrix elements of the Hamiltonian are pre-computed and stored in memory or on disk. Using the known importance weights as initial pivots, one can obtain convergence of a single target states after typically 10 iterations. Eventually, we obtain energy eigenvalues and amplitudes of the target states. Since the eigenstates are---at no additional cost---given in a simple shell-model representation, we can easily use them for subsequent computation of various expectation values and density distributions or form-factors.
 
The time-consuming parts to the code, i.e. the construction of the importance-truncated space and the computation of the Hamilton matrix, can be easily parallelized with practically no communication overhead and perfect scaling. We use a hybrid OpenMP plus MPI parallelization strategy to make optimal use of the memory resources of modern multi-core architectures. As compare to a typical full NCSM, the particle numbers and model space sizes in the importance-truncated NCSM are not limited by the available memory. Larger model spaces or particle numbers only require more CPU-time for the construction of the importance-truncated model space.

\subsection{Iterative Construction of Model-Space: IT-NCSM($i$) vs. IT-NCSM(seq)}
\label{sec:itncsm_importanceupdate}

We can use the universal update scheme described in Sec.~\ref{sec:its_modelspace} for the iterative construction of the importance truncated $N_{\max}\hbar\Omega$ space for any given $N_{\max}$. For targeting the ground state, we would start with a $0\hbar\Omega$ eigenstate as the initial reference state---for a closed-shell nucleus this is just the independent-particle shell-model determinant. In a first iteration the importance update is used to generate all relevant $1p1h$ and $2p2h$ excitations within the $N_{\max}\hbar\Omega$ space under consideration. Using the eigenstate in this importance-truncated space as reference state, a second iteration will give access to all basis states up to the $4p4h$ level with respect to the initial $0\hbar\Omega$ state. Typically two or three iterations of the importance update cycle are sufficient to obtain convergence, i.e. a result which is not changed anymore by another importance update. In the following we will identify those calculations with the label IT-NCSM($i$), where $i$ indicates the number of iterations. 

However, for the $N_{\max}\hbar\Omega$ space of the NCSM there exists a more efficient alternative. Typically we are interested in a sequence of calculations for growing $N_{\max}$ in order to assess the convergence behavior with increasing model-space size. We can combine this sequential increase of $N_{\max}$ with the importance update in an elegant way. Assume we start with a complete NCSM calculation in a $0\hbar\Omega$ or $2\hbar\Omega$ space. Using the eigenstate obtained in this small, say $2\hbar\Omega$ space as reference state we construct the importance truncated $4\hbar\Omega$ space and solve the eigenvalue problem again. The resulting eigenstate then defines the reference state for the construction of the importance truncated $6\hbar\Omega$ space, and so on. We will identify calculations based on this sequential update scheme by IT-NCSM(seq) in the following.

This sequential scheme has an important conceptual advantage: The maximum $npnh$ excitation with respect to the $0\hbar\Omega$ space that is contained in an $N_{\max}\hbar\Omega$ space is of order $n=N_{\max}$. Therefore in each step of sequence $N_{\max}=0,2,4,6,...$ the maximum $npnh$-order increases by 2 and a single importance update at each step is sufficient to access all $npnh$ orders that can appear. Thus, the sequential update scheme recovers the complete $N_{\max}\hbar\Omega$ model space in the limit $(\kappa_{\min},C_{\min})\to 0$ and does not impose any explicit limitation regarding the $npnh$-content of the space. We need to apply the importance update only once for each value of $N_{\max}$, in the iterative scheme we would need $i=N_{\max}/2$ iterations to formally achieve this. We will apply and compare both schemes in Sec. \ref{sec:app}.

\subsection{Importance Measure}
\label{sec:itncsm_importancemeasure}

\begin{figure}
\includegraphics[width=1\columnwidth]{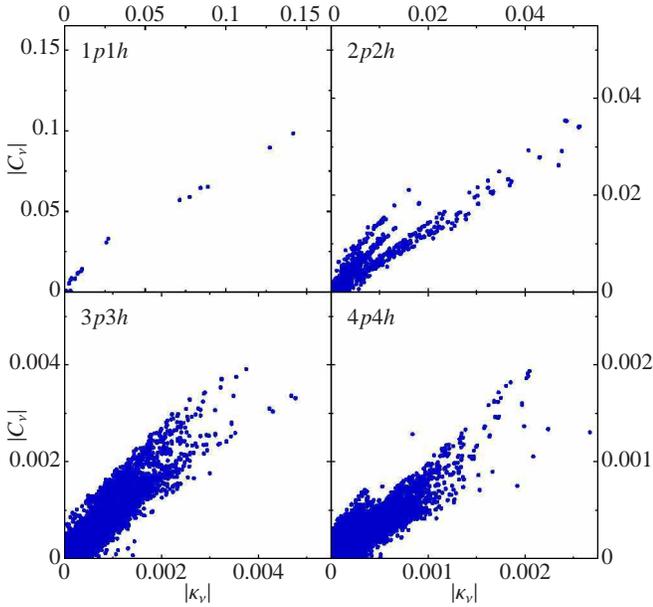}
\caption{(color online) Correlation between the importance measure $\kappa_{\nu}$ and the amplitude $C_{\nu}$ obtained by solving the eigenvalue problem in an IT-NCSM(2) calculation of \elem{O}{16} with $N_{\max}=8$ and $\hbar\Omega=22$ MeV. The panels correspond to the different $npnh$-orders as indicated. }
\label{fig:itncsm_coeffcorr}
\end{figure}

As a first test of the reliability of the importance measure, we can compare the perturbative estimate $\kappa_{\nu}$ for the amplitude of a given basis state $\ket{\Phi_{\nu}}$ with the amplitude $C_{\nu}$ resulting from the diagonalization. Whereas the \emph{a priori} importance measure $\kappa_{\nu}$ only includes the coupling to the states from the reference space, the \emph{a posteriori} amplitudes $C_{\nu}$ are affected by the mutual coupling of all states. Nonetheless, the $\kappa_{\nu}$ provides a reasonable estimate for the amplitudes $C_{\nu}$ which is sufficient to identify the important basis states. 

This is illustrated in Fig.~\ref{fig:itncsm_coeffcorr} for an importance-truncated NCSM calculation for \elem{O}{16} in an $8\hbar\Omega$ space using two iterations of the importance-update of the model space for an importance threshold  $\kappa_{\min} = 5\times10^{-5}$. The correlation plots relate the importance measure $\kappa_{\nu}$ of the individual basis states with the corresponding amplitudes $C_{\nu}$ in the final eigenstates. There is a clear correlation between the two quantities which is sufficient to predict which basis states are important for an adequate representation of the final eigenstate. The scattering around the diagonal reflects all couplings that are not accounted for in the lowest-order perturbative estimate.

\begin{figure}
\includegraphics[width=0.7\columnwidth]{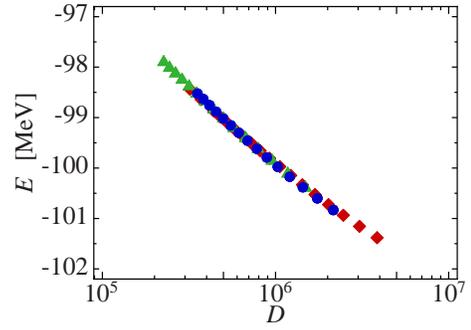}
\caption{(color online) Ground-state energy of \elem{O}{16} obtained in IT-NCSM(2) calculations for $N_{\max}=8$ and $\hbar\Omega=22$ MeV as function of the dimension $D$ of the importance-truncated space. The different symbols correspond to different definitions of the importance measure: the state-based importance measure $\kappa_{\nu}$ (\symbolcircle[FGBlue]), the energy-based importance measure $\chi_{\nu}$ (\symboldiamond[FGRed]), and the state-based importance measure $\kappa^{\text{EN}}_{\nu}$ defined in an Epstein-Nesbet partitioning (\symboltriangle[FGGreen]).}
\label{fig:itncsm_kapvschi}
\end{figure}

As mentioned in Sec.~\ref{sec:its_importancemeasure} there are other options to define an importance measure in the framework of multiconfigurational perturbation theory. A natural alternative to the state-based importance measure $\kappa_{\nu}$ is the energy-based importance measure $\chi_{\nu}$ defined in Eq.~\eqref{eq:its_importancemeasureenergy}. One could also consider an Epstein-Nesbet partitioning as discussed in Sec.~\ref{sec:its_perturbationtheory} to set up the perturbative corrections and define a state-based importance measure $\kappa^{\text{EN}}_{\nu}$. 

In order to assess the efficiency of the three measures we perform a series of calculations with different values of the respective importance thresholds and plot the energy eigenvalue versus the dimension of the importance truncated space as a parametric curve spanned by the importance thresholds $\kappa_{\min}$, $\chi_{\min}$, and $\kappa^{\text{EN}}_{\min}$, respectively. Since the whole approach is variational, the measure which leads to the lowest ground-state energy for a given dimension $D$ of the importance-truncated space is most efficient in selecting the $D$ most important basis states. 

An example of this analysis is shown in Fig. \ref{fig:itncsm_kapvschi}, again for the ground state of \elem{O}{16} in an $8\hbar\Omega$ space. In all cases the NCSM ground state in a complete $2\hbar\Omega$ space was used as reference state for the construction of the importance-truncated space. The points obtained with all three definitions of the importance measure essentially fall onto the same line, i.e. all measures are able to identify the most important configurations with the same efficiency. We therefore use the conceptually and computationally simplest importance measure, the state-based measure $\kappa_{\nu}$ of Eq. \eqref{eq:its_importancemeasure} in all following investigations.

\subsection{Threshold Dependence \& Extrapolation}
\label{sec:itncsm_thresholdextrapol}

The variation of the threshold $\kappa_{\min}$ is an important probe for the quality of the importance truncation and the basis for an extrapolation to vanishing threshold $\kappa_{\min}\to0$ as it will be used later on. To this end, all IT-NCSM calculations are performed for a sequence of different values for $\kappa_{\min}$. For each threshold value the importance truncated space is different and the eigenvalue problem has to be solved again. However, this can be done at small computational cost. The importance truncated space and the Hamilton matrix are initially determined for the smallest $\kappa_{\min}$. After the solution of the eigenvalue problem for this threshold, all basis states that are not part of space for the next-larger importance threshold and the corresponding matrix elements are removed, and the eigenvalue problem is solved again. Hence, the time consuming construction of the importance truncated space and the computation of the Hamilton matrix is done only once for a whole threshold sequence.

\begin{figure}
\includegraphics[width=0.7\columnwidth]{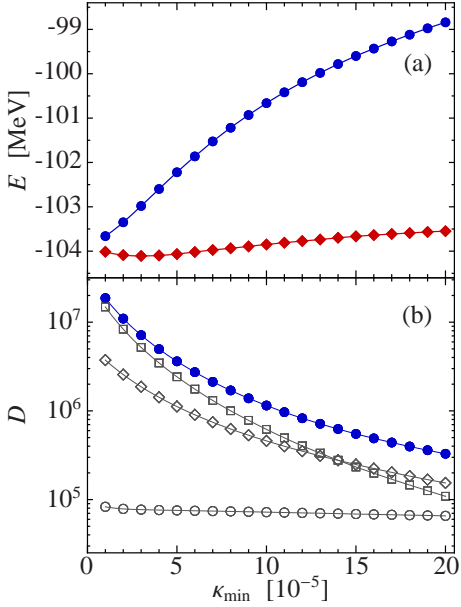}
\caption{(color online) Threshold dependence of the energies and the model space dimension for a IT-NCSM(2) calculation of \elem{O}{16} with $N_{\max}=8$ and $\hbar\Omega=22$ MeV. (a) Energy eigenvalues and as function of $\kappa_{\min}$ without (\symbolcircle[FGBlue]) and with (\symboldiamond[FGRed]) perturbative correction for the excluded configurations. (b) Total dimension of the importance truncated space (\symbolcircle[FGBlue]) as well as the number of $2p2h$ (\symbolcircleopen[FGDarkGray]), $3p3h$ (\symboldiamondopen[FGDarkGray]), and $4p4h$-configurations (\symbolboxopen[FGDarkGray]) with varying $\kappa_{\min}$.}
\label{fig:itncsm_thesholddep}
\end{figure}

The dependence of the energy and of the model-space dimension on the importance threshold $\kappa_{\min}$ in IT-NCSM(2) calculations for \elem{O}{16} with different $N_{\max}$ is illustrated in Fig.~\ref{fig:itncsm_thesholddep}. The energy eigenvalue $E(\kappa_{\min})$ obtained in the importance truncated space decreases monotonically with decreasing $\kappa_{\min}$ as expected from the variational principle and the Hylleraas-Undheim theorem. At the same time, the dimension of the importance truncated space increases exponentially with decreasing $\kappa_{\min}$. The number of configurations of higher $npnh$-order in particular grows rapidly as the threshold is lowered. This behavior reflects the mechanism behind the importance truncation scheme: The configurations which are most important for the description of the target state have large $\kappa_{\nu}$ and are included already for large thresholds. With decreasing threshold $\kappa_{\min}$ basis states of lesser importance are successively included. Their number increases dramatically but the effect on the state and the energy remains moderate, facilitating approximations to estimate their effect on the energy without including them explicitly in the model space.
 
The simplest approximate way to account for the excluded basis states is the \emph{a posteriori} energy correction $\Delta_{\text{excl}}(\kappa_{\min})$ given by Eq. \eqref{eq:its_pertcorrectionexcluded} on the basis of the second-order MCPT contribution. The corrected energies $E(\kappa_{\min})+\Delta_{\text{excl}}(\kappa_{\min})$ are also depicted in Fig.~\ref{fig:itncsm_thesholddep}(a). Although $\Delta_{\text{excl}}(\kappa_{\min})$ provides only a rough estimate for the contribution of excluded states, the $\kappa_{\min}$-dependence of the corrected energy is much weaker than the dependence of the uncorrected eigenvalues $E(\kappa_{\min})$---if the correction were exact we would expect the corrected energies to be independent of $\kappa_{\min}$. In many cases the corrected energy at a single value of $\kappa_{\min}$ can already serve as a good approximation for the full results in the limit $\kappa_{\min}\to0$.

\begin{figure}
\includegraphics[width=0.7\columnwidth]{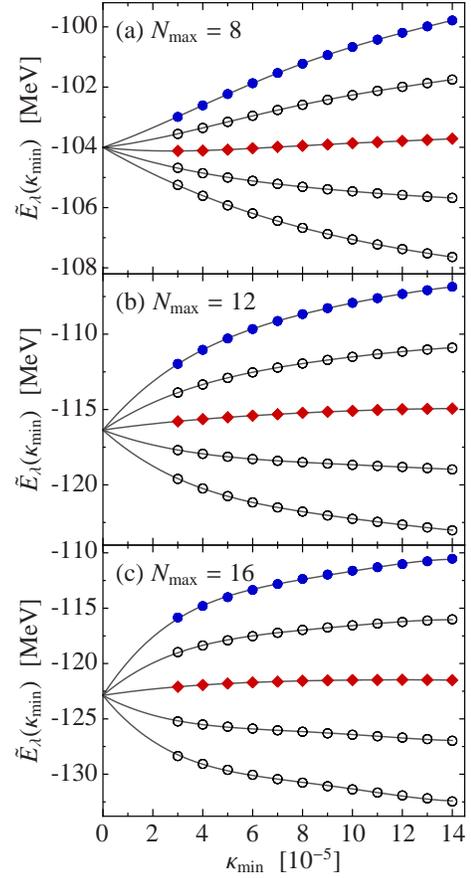}
\caption{(color online) Threshold extrapolation of the ground-state energy of \elem{O}{16} ($\hbar\Omega=22$ MeV) obtained in IT-NCSM(2) for different $N_{\max}$. Shown are the perturbatively corrected energies $E_{\lambda}(\kappa_{\min})$ as function of $\kappa_{\min}$ for $\lambda=0, 0.5, 1, 1.5,$ and $2$ (data sets from top to bottom within each panel). For $\lambda=0$ (\symbolcircle[FGBlue]) the original energy eigenvalue $E(\kappa_{\min})$ is recovered, for $\lambda=1$ (\symboldiamond[FGRed]) we obtain the perturbatively corrected energy $E(\kappa_{\min})+\Delta_{\text{excl}}(\kappa_{\min})$. The lines show the results of a simultaneous constrained fit for all data sets using 4th order polynomials (see text).}
\label{fig:itncsm_thresholdextrapol}
\end{figure}

A more reliable way to recover the contribution of excluded configurations is an \emph{a posteriori} extrapolation of the energies to vanishing importance threshold. Due to the smooth and monotonic behavior of the energies $E(\kappa_{\min})$ one can attempt a direct numerical extrapolation $\kappa_{\min}\to0$ as done in Ref.~\cite{RoNa07}. Since the general shape of the $E(\kappa_{\min})$ curve varies, we will generally use polynomials in $\kappa_{\min}$ fitted to a sufficiently large number of different threshold values for the extrapolation. Instead of $E(\kappa_{\min})$ one can extrapolate the perturbatively corrected energy, $E(\kappa_{\min})+\Delta_{\text{excl}}(\kappa_{\min})$, which shows a weaker threshold dependence than the eigenvalues and, therefore, allows for a more stable extrapolation. The extrapolation can be stabilized further by performing a simultaneous fit of $E(\kappa_{\min})$ and $E(\kappa_{\min})+\Delta_{\text{excl}}(\kappa_{\min})$. Since the perturbative correction $\Delta_{\text{excl}}(\kappa_{\min})$ has to vanish in the limit $\kappa_{\min}\to0$ both extrapolations should formally give the same value at $\kappa_{\min}=0$, independent of the absolute quality of the perturbative estimate. The formal property $E(0)=E(0)+\Delta_{\text{excl}}(0)$ is used as a constraint in the simultaneous fit and reduces the uncertainties of the threshold extrapolation significantly. 

One can even go one step further and define a family of energy curves $\tilde{E}_{\lambda}(\kappa_{\min}) = E(\kappa_{\min})+\lambda \Delta_{\text{excl}}(\kappa_{\min})$ with a control parameter $\lambda$. Independent of the choice of $\lambda$ the formal property $\tilde{E}_{\lambda}(0) = E(0)$ holds. Using this as a constraint in a simultaneous $\chi^2$-fit of a set of curves for several values of $\lambda$ provides very robust extrapolation results. This technique has been pioneered by Buenker and Peyerimhoff in the early applications of configuration-selecting CI approaches in quantum chemistry \cite{BuPe75}. It solely relies on the fact that the correction $\Delta_{\text{excl}}(\kappa_{\min})$ is a monotonous function which goes to zero (smoothly) as $\kappa_{\min}\to0$. 

Examples for this type of threshold extrapolation in the case of IT-NCSM(2) calculations for \elem{O}{16} in different $N_{\max}\hbar\Omega$ model spaces are presented in Fig. \ref{fig:itncsm_thresholdextrapol}.
The starting point are the energies $E(\kappa_{\min})$ and perturbative corrections $\Delta_{\text{excl}}(\kappa_{\min})$ obtained for a sequence of importance thresholds in the range from $\kappa_{\min} = 3\times10^{-5}$ to $14\times10^{-5}$. Using this input we construct data sets for the corrected energies $\tilde{E}_{\lambda}(\kappa_{\min})$ for $\lambda=0, 0.5, 1, 1.5,$ and $2$ and simultaneously fit each of the sets by a 4th order polynomial under the constraint that all curves meet at $\kappa_{\min}=0$. The individual data sets and the polynomial fits are shown in Fig.~\ref{fig:itncsm_thresholdextrapol}. It is evident that this extrapolation scheme is most stable if the curves approach the common $\tilde{E}_{\lambda}(0)$ value more or less symmetrically. This is the reason for the particular set of $\lambda$-values adopted here. 

We employ the following threshold extrapolation protocol for the applications presented in Sec.~\ref{sec:app}. Using a sequence of 12 equidistant threshold values in the range $\kappa_{\min} = 3\times10^{-5}$ to $14\times10^{-5}$ we perform a constrained simultaneous fit of the corrected energies $E_{\lambda}(\kappa_{\min})$ for a sequence of at least 5 different $\lambda$-values using low-order polynomials. The set of $\lambda$-parameters is chosen such that the common point of all fit curves at $\kappa_{\min}=0$, which gives the final threshold-extrapolated energy, is approached symmetrically. In order to assess the uncertainty of the extrapolation, we drop the smallest and the largest value, respectively, from the $\lambda$-sequence and perform the simultaneous fit for the remaining data sets. The variance of this set of extrapolations defines an uncertainty interval for the threshold extrapolated energy. 

Exceptions are very light nuclei, e.g. \elem{He}{4}, where a direct extrapolation of the energy eigenvalue $E(\kappa_{\min})$ without using the perturbative correction $\Delta_{\text{excl}}(\kappa_{\min})$ provides a more stable result. The reason is the $\kappa_{\min}$-dependence of $\Delta_{\text{excl}}(\kappa_{\min})$, which in very small spaces shows structures that interfere with the polynomial extrapolation.

\section{Applications \& Benchmarks: Magic Nuclei}
\label{sec:app}

We employ the IT-NCSM now for the series of calculations for the $0^+$ ground state energies of various closed and open shell nuclei in the p-shell. The aim is to compare the results to the full NCSM in different cases in order to demonstrate the robustness of the importance truncation scheme. All full NCSM calculations presented in the following were performed with the \textsc{Antoine} code \cite{CaNo99}.

\subsection{Helium-4}
\label{sec:app_He4}

As the simplest benchmark we study the ground-state energy of \elem{He}{4} using the $V_{\UCOM}$ interaction. In this case full NCSM calculations can be performed up to very large $N_{\max}\hbar\Omega$ spaces such that convergence is observed. Furthermore, other few-body methods, e.g. the hyperspherical harmonics basis expansion \cite{Bacc07}, have been employed and yield an independent reference value for the ground-state energy.

First we consider the simple iterative scheme IT-NCSM($i$) for the construction of the importance truncated model space. For fixed $N_{\max}$ we perform up to three iterations of the importance update starting with the Slater determinant of the  independent-particle model as initial reference. In each iteration we solve the eigenvalue problem for a sequence of importance thresholds in the range $\kappa_{\min}=3\times10^{-5}$ to $14\times10^{-5}$ and extrapolate the eigenvalues $E(\kappa_{\min})$ to the limit of vanishing threshold $\kappa_{\min}\to0$ as discussed in Sec.~\ref{sec:itncsm_thresholdextrapol}. For very light nuclei like \elem{He}{4} the direct extrapolation of $E(\kappa_{\min})$ without perturbative corrections for excluded configuration provides the most stable results. For the definition of the reference state for the next iteration a reference threshold $C_{\min}=5\times10^{-4}$ is used.

\begin{figure}
\includegraphics[width=0.75\columnwidth]{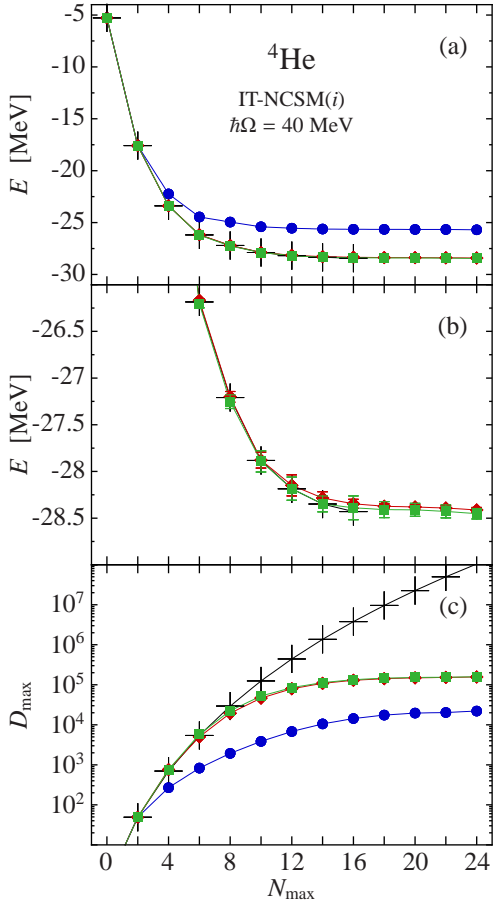}
\caption{(color online) Ground-state energy and model-space dimension as function of $N_{\max}$ for \elem{He}{4} obtained within the IT-NCSM($i$) scheme for $i=1$ (\symbolcircle[FGBlue]), $i=2$ (\symboldiamond[FGRed]), and $i=3$ iterations (\symbolbox[FGGreen]) using the $\VO_{\UCOM}$ interaction for $\hbar\Omega=40$ MeV. Panels (a) and (b) show the ground-state energies on different scales, including the uncertainty estimates for the threshold extrapolation. Panel (c) depicts the maximum dimension of the importance-truncated model space. For comparison the results of full NCSM calculations for the same Hamiltonian are included ($+$).}
\label{fig:app_He4_iterative}
\end{figure}

The threshold-extrapolated ground-state energies and the dimensions of the maximum importance-truncated model spaces as function of $N_{\max}$ are depicted in Fig.~\ref{fig:app_He4_iterative}. The convergence with respect to the importance updates of the model space is very fast. After two iterations, i.e. at the IT-NCSM(2) level, we already obtain stable results which is within 100 keV of the full NCSM result. The third iteration only lowers the ground-state energy a little further bringing it into excellent agreement with the full NCSM, as seen in Fig.~\ref{fig:app_He4_iterative}(b). In the case of \elem{He}{4} this convergence pattern may be expected. After two iterations the importance truncated space contains up to $4p4h$ excitations, i.e., the full model space can be generated in the limit of vanishing thresholds. The minimal change in the third iteration is due to a relaxation of the importance truncated space, i.e., through the reassessment of the importance of all basis states with respect to a new reference state, which includes all possible $npnh$-orders, the importance truncated space is better adapted . Further importance updates do not change the resulting energies anymore.  

The agreement with the full NCSM demonstrates the efficiency of the importance measure and the reliability of the threshold extrapolation. The dimension $D_{\max}$ of the largest model space considered for the threshold extrapolation is up to two orders of magnitude smaller than the dimension of the full NCSM space, as illustrated in Fig.~\ref{fig:app_He4_iterative}(c). Note that the full NCSM dimension is obtained by exploiting all relevant symmetries, including parity and time-reversal, to reduce the dimension of the eigenvalue problem---it corresponds to the `effective dimension' used by the \textsc{Antoine} code. Thus this substantial reduction of the model-space dimension by the importance truncation goes beyond generic symmetries and really exploits the specific properties of the Hamiltonian. 

\begin{figure}
\includegraphics[width=0.75\columnwidth]{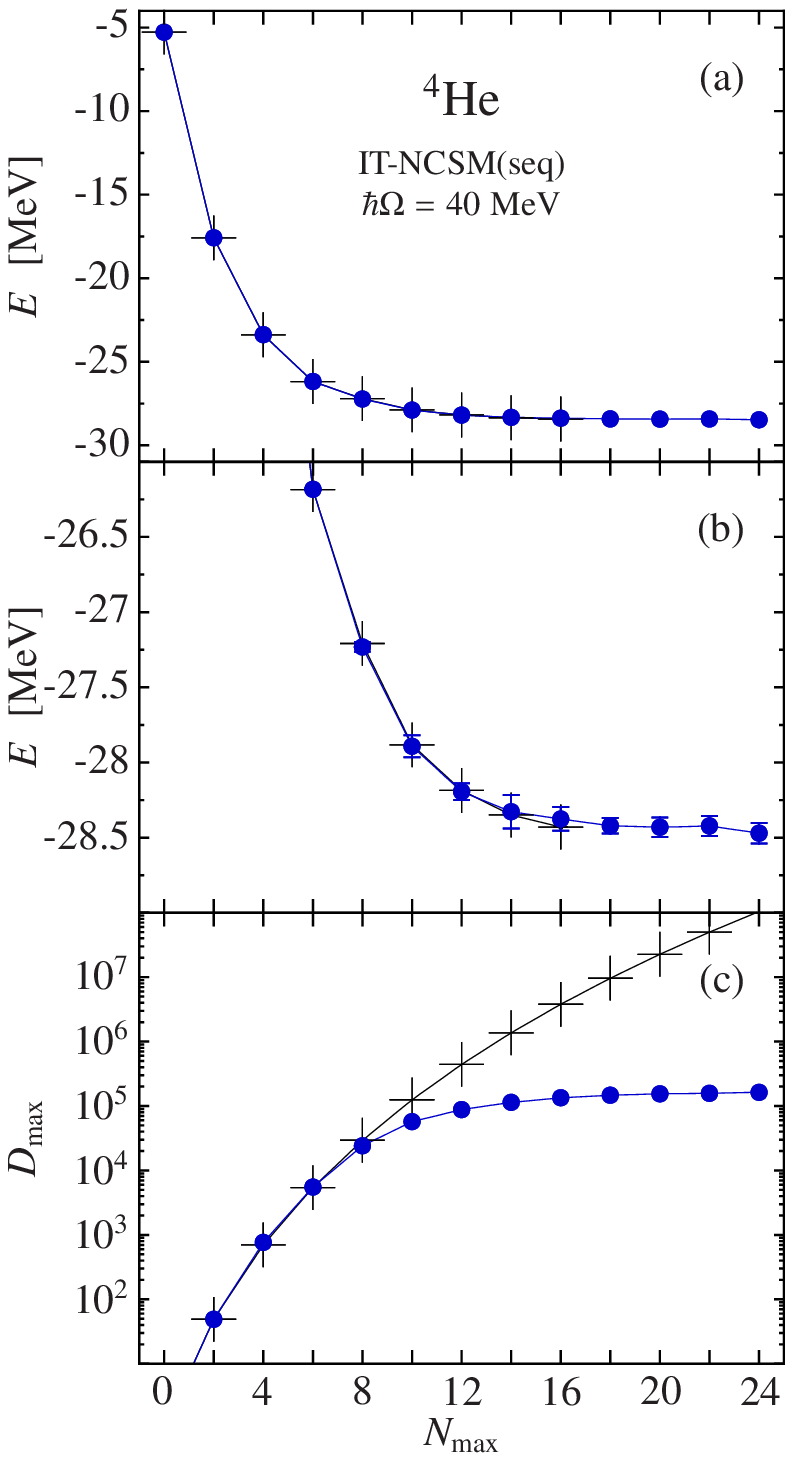}
\caption{(color online) Ground-state energy, (a) and (b), and model-space dimension (c) as function of $N_{\max}$ for \elem{He}{4} obtained within the IT-NCSM(seq) scheme (\symbolcircle[FGBlue]) using the $\VO_{\UCOM}$ interaction for $\hbar\Omega=40$ MeV. For comparison the results of full NCSM calculations with the same Hamiltonian are included ($+$).}
\label{fig:app_He4_sequential}
\end{figure}

As an alternative to the simple iterative model-space update at fixed $N_{\max}$ we can perform these calculations using the sequential model-space update IT-NCSM(seq) proposed in Sec.~\ref{sec:itncsm_importanceupdate}. Starting from the $0\hbar\Omega$ space we use the importance measure to construct an importance truncated $2\hbar\Omega$ space. This is used as reference space to construct the importance truncated $4\hbar\Omega$ space, and so on. As before we use a reference threshold of $C_{\min}=5\times10^{-4}$ and a sequence of importance thresholds starting from $\kappa_{\min}=3\times10^{-5}$. The results for the ground-state energies of \elem{He}{4} are summarized in Fig.~\ref{fig:app_He4_sequential} and compared to the full NCSM. The IT-NCSM(seq) scheme leads to the same excellent agreement with the full NCSM as the IT-NCSM(3). However, the IT-NCSM(seq) is computationally more efficient, since only one importance update is needed for each value of $N_{\max}$. 

\begin{figure}
\includegraphics[width=0.8\columnwidth]{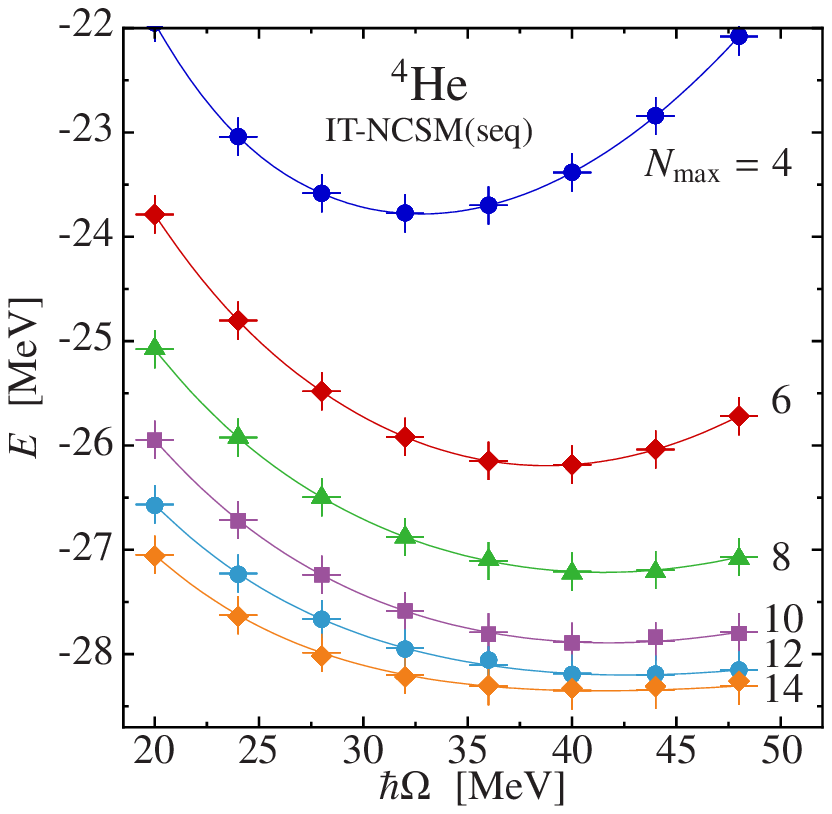}
\caption{(color online) Ground-state energies of \elem{He}{4} obtained for the $\VO_{\UCOM}$ interaction as function of the oscillator frequency $\hbar\Omega$ for different $N_{\max}\hbar\Omega$ model spaces. Shown are the results of IT-NCSM(seq) calculations (full symbols) in comparison to full NCSM calculations (crosses).}
\label{fig:app_He4_sequential_hwdep}
\end{figure}

The dependence of the ground-state energy obtained in the IT-NCSM(seq) on the oscillator parameter $\hbar\Omega$ is depicted in Fig.~\ref{fig:app_He4_sequential_hwdep}. The comparison with the full NCSM results shows that the excellent agreement persists for all frequencies $\hbar\Omega$. The particular oscillator frequency $\hbar\Omega=40$ MeV used in Figs.~\ref{fig:app_He4_iterative} and \ref{fig:app_He4_sequential} corresponds to the minimum for the larger space. 

In order to compare our results with other many-body methods and with experiment, we perform an exponential extrapolation of the IT-NCSM(seq) energies for $\hbar\Omega=40$ MeV. Since the calculations are practically converged with respect to $N_{\max}$ the main purpose of the extrapolation is to smooth out the fluctuations due to the uncertainties of the threshold extrapolation. Using the five data points from $N_{\max}=16$ to $24$ we obtain a \elem{He}{4} ground-state energy of $-28.52(10)$ MeV. This is in excellent agreement with the value of $-28.57$ MeV that was obtained previously in the framework of the hyperspherical harmonics approach using the same $V_{\UCOM}$ interaction \cite{Bacc07}. The comparison to the experimental binding energy of $-28.29$ MeV only reveals the rough nature of the adjustment of the UCOM tensor correlator range $I_{\vartheta}$ that was used in Ref. \cite{RoHe05} to fix the $V_{\UCOM}$ interaction. In principle one could select $I_{\vartheta}$ such that the experimental \elem{He}{4} binding energy is reproduced exactly.

\subsection{Oxygen-16}
\label{sec:app_O16}

The ground state of \elem{O}{16} poses a more challenging problem. At present, full NCSM calculations can be done routinely for spaces up to $N_{\max}=8$ with an effective dimension of almost $0.6\times10^9$. For $N_{\max}=10$ and $12$ the effective dimension grows to $1.4\times10^{10}$ and $2.4\times10^{11}$, respectively, which is clearly beyond the reach of present NCSM codes. The importance truncation is crucial in this domain and enables us to treat model spaces of up to $N_{\max}=22$ and beyond. This limit is set by the available two-body matrix elements and not by the IT-NCSM calculation itself. 

\begin{figure}
\includegraphics[width=0.75\columnwidth]{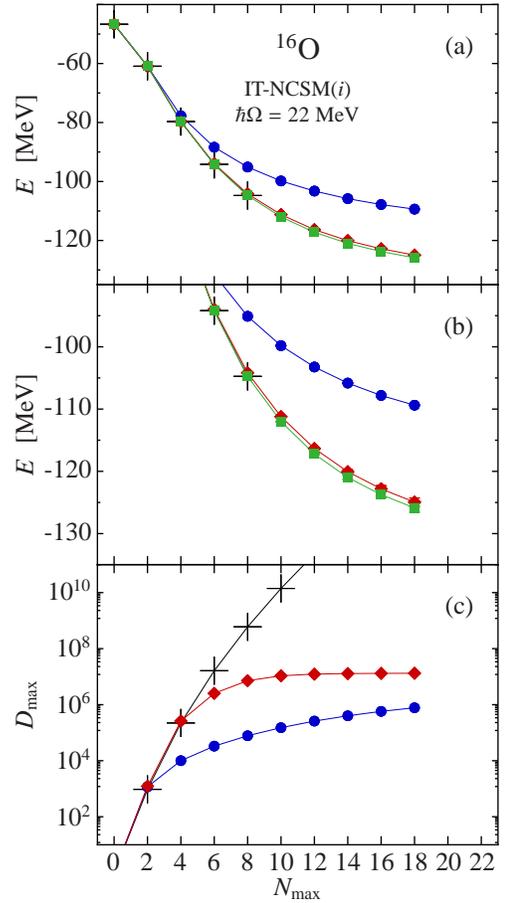}
\caption{(color online) Ground-state energy and model-space dimension as function of $N_{\max}$ for \elem{O}{16} obtained within the IT-NCSM($i$) scheme for $i=1$ (\symbolcircle[FGBlue]) and $i=2$ (\symboldiamond[FGRed]) iterations using the $\VO_{\UCOM}$ interaction for $\hbar\Omega=22$ MeV. In addition the IT-NCSM(2) energies after inclusion of the multi-reference Davidson correction are shown (\symbolbox[FGGreen]).  Panels (a) and (b) show the ground-state energies on different scales, including the uncertainty estimates for the threshold extrapolation. Panel (c) depicts the maximum dimension of the importance-truncated model space. For comparison the results of full NCSM calculations for the same Hamiltonian are included ($+$).}
\label{fig:app_O16_iterative}
\end{figure}

\begin{table}
\caption{Ground-state energies (in units of MeV) for \elem{O}{16} obtained for the $\VO_{\UCOM}$ interaction at $\hbar\Omega=22$MeV with different levels of the IT-NCSM. For the IT-NCSM($i$) results for $i=1$  and $2$ iterations are shown. Furthermore IT-NCSM(2) results with the MRD correction \eqref{eq:its_mrdavidsoncorrection} and the perturbative correction \eqref{eq:its_pertcorrectionnextiteration} for the effect of the next iteration are reported. For the IT-NCSM(seq) two different reference thresholds have been used: (a) $C_{\min}=5\times10^{-4}$ and (b) $C_{\min}=3\times10^{-4}$. Numbers in parentheses are uncertainty estimates for the threshold extrapolation.}
\label{tab:app_O16_summary}
\begin{ruledtabular}\small
\begin{tabular}{l d d d}
$N_{\max}$ & 8 & 12 & 16 \\
\hline
$E_0$      & -46.69 & -46.69  & -46.69 \\
IT-NCSM(1) & -95.10(2) & -103.24(2) & -107.81(2) \\
IT-NCSM(2) & -104.18(15) & -116.32(15) & -122.81(50) \\
IT-NCSM(2)+MRD &  -104.75(15) & -117.22(15) & -123.75(50) \\
IT-NCSM(2)+PT(3) & -104.81(15) & -117.62(15) & - \\
\hline
IT-NCSM(seq) - (a) & -104.49(10) & -116.86(25) & -123.14(70) \\
IT-NCSM(seq) - (b) & -104.43(10) & -117.12(25) & -123.45(70) \\
\hline
full NCSM  & -104.75 & - & - \\
\end{tabular}
\end{ruledtabular}
\end{table}

As for \elem{He}{4}, we first consider the simple iterative IT-NCSM($i$) scheme using up to two iterations for each $N_{\max}$ to construct the importance truncated model space. We use a set of 12 equidistant importance thresholds in the range $\kappa_{\min}=3\times10^{-5}$ to $14\times10^{-5}$ as input for the simultaneous threshold extrapolation as discussed in Sec. \ref{sec:itncsm_thresholdextrapol}. The reference threshold is set to $C_{\min}=5\times10^{-4}$. 

A summary of the IT-NCSM($i$) results for the ground-state energies of \elem{O}{16} up to $N_{\max}=18$ is presented in Fig. \ref{fig:app_O16_iterative}, selected numerical values are given in Tab. \ref{tab:app_O16_summary}. As for the much lighter nucleus \elem{He}{4} the convergence of the iterative importance updates is excellent. Already after two iterations, i.e. for IT-NCSM(2), the full NCSM energies up to $N_{\max}=8$ are produced to an absolute accuracy of better than $600$ keV. 

Instead of performing a third iteration explicitly, we can use computationally simpler estimates for the small correction resulting from $5p5h$ and $6p6h$ configuration that are not present in the IT-NCSM(2) model space. As discussed in Sec. \ref{sec:its_corrections}, the simplest \emph{a posteriori} correction is the multi-reference Davidson correction (MRD) given by Eq. \eqref{eq:its_mrdavidsoncorrection} since it does not require any additional computation beyond IT-NCSM(2). The MRD corrected IT-NCSM(2) energies are also shown in Figs. \ref{fig:app_O16_iterative}(a) and (b). The contribution of the MRD correction grows slightly with $N_{\max}$ and reaches about $1$ MeV for $N_{\max}=18$. As seen from Tab.~\ref{tab:app_O16_summary}, the IT-NCSM(2)+MRD energy is in excellent agreement with the full NCSM. A computationally more demanding \emph{a posteriori} correction based on the explicit calculation of the second-order energy contribution on top of the IT-NCSM(2) eigenstate as defined by Eq. \eqref{eq:its_pertcorrectionnextiteration} yields very similar results. The IT-NCSM(2)+PT(3) energies shown in Tab. \ref{tab:app_O16_summary} agree very well with both, IT-NCSM(2)+MRD and full NCSM.

The good agreement with the full NCSM energies is yet another indication of the efficiency of the importance truncation scheme in selecting the relevant configurations and of the reliability of the threshold extrapolation for recovering the contribution of excluded configurations. The importance truncated space is substantially smaller than full NCSM space as seen in Fig. \ref{fig:app_O16_iterative}(c). For $N_{\max}=8$ the importance truncation reduces the dimension by two orders of magnitude, for $N_{\max}=12$ already by four orders of magnitude. This dramatic reduction allows us to go to much larger values of $N_{\max}$ than ever possible in the full NCSM.

\begin{figure}
\includegraphics[width=0.75\columnwidth]{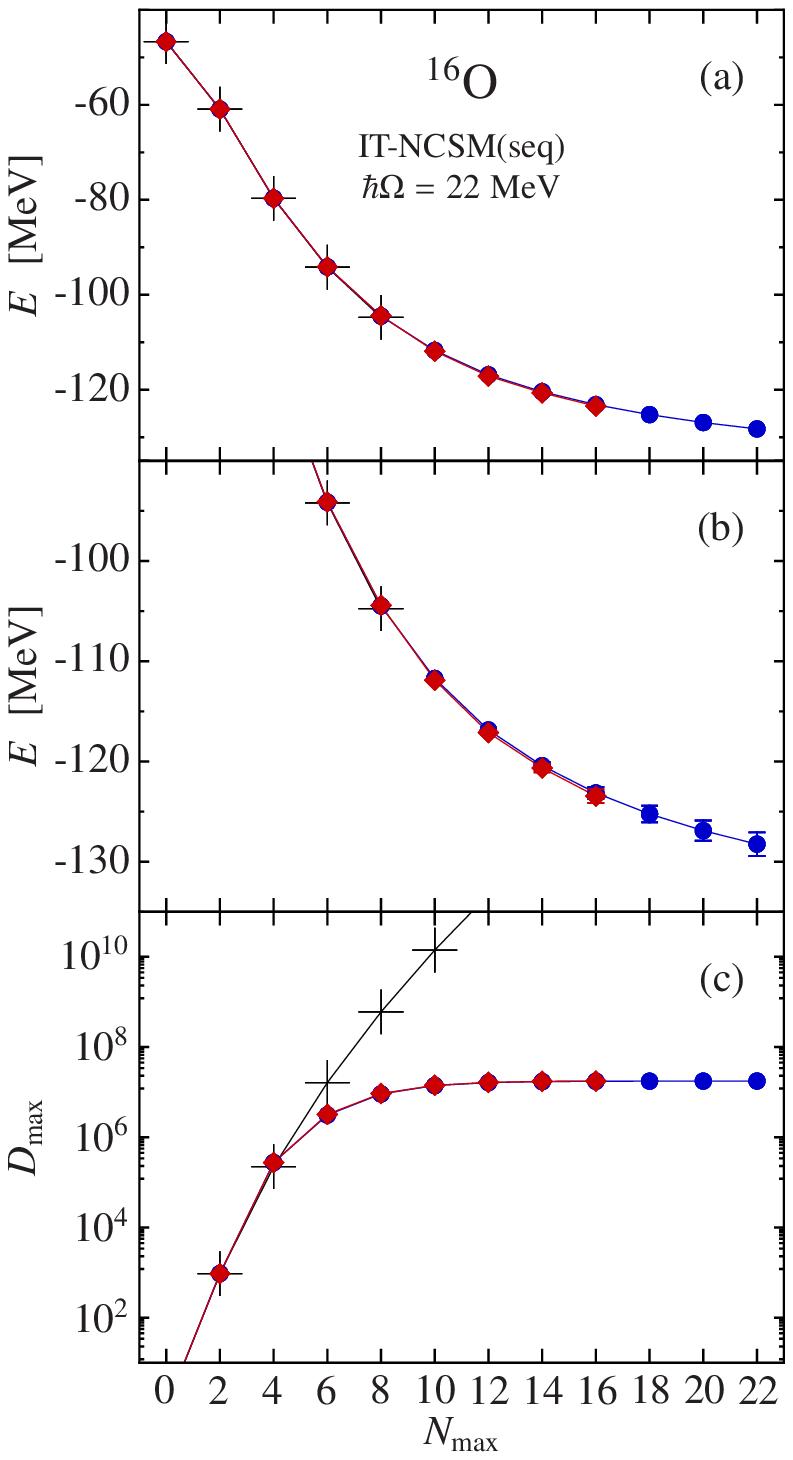}
\caption{(color online) Ground-state energy, (a) and (b), and model-space dimension (c) as function of $N_{\max}$ for \elem{O}{16} obtained within the IT-NCSM(seq) scheme  using the $\VO_{\UCOM}$ interaction for $\hbar\Omega=22$ MeV. Two different values of the parent threshold were used: $C_{\min}=5\times10^{-4}$ (\symbolcircle[FGBlue]) and  $C_{\min}=3\times10^{-4}$ (\symboldiamond[FGRed]). For comparison the results of full NCSM calculations with the same Hamiltonian are included ($+$).}
\label{fig:app_O16_sequential}
\end{figure}
 
We can improve the efficiency even further by using the sequential IT-NCSM(seq) scheme, which requires only one importance update for each value of $N_{\max}$ since it uses a reference state constructed from the eigenstate in the $N_{\max}-2$ space. In this way all $npnh$-states that are possible in a given $N_{\max}\hbar\Omega$ space are generated in the limit $(\kappa_{\min},C_{\min})\to0$. 

The results of IT-NCSM(seq) calculations for the ground-state energy of \elem{O}{16} for $\hbar\Omega=22$ MeV and the sequence of $N_{\max}$ values starting from $N_{\max}=0$ up to $N_{\max}=22$ are presented in Fig. \ref{fig:app_O16_sequential}. We study two different values of $C_{\min}$, the threshold used in the definition of the reference state, since this is the only parameter left after the $\kappa_{\min} \to 0$ extrapolation. The set of $\kappa_{\min}$ values used for the threshold extrapolation is the same as before.

We observe an excellent agreement with the full NCSM and with the IT-NCSM(2) of Fig. \ref{fig:app_O16_iterative}. The numerical results in Tab. \ref{tab:app_O16_summary} reveal that the IT-NCSM(seq) energies are slightly but systematically below the IT-NCSM(2) results. This is due to the presence of $5p5h$ and $6p6h$ configurations in the model space of the IT-NCSM(seq), which are excluded from the IT-NCSM(2) space. States beyond the $6p6h$ level are suppressed by the importance truncation, i.e. they do not have importance measures above the smallest threshold $\kappa_{\min}=3\times10^{-5}$ used in this calculation.
The IT-NCSM(seq) calculations for the two different reference thresholds $C_{\max}$ agree within the uncertainties of the $\kappa_{\min}$ extrapolation, which indicates that the values chosen here are sufficiently small to capture all relevant components of the reference state.

\begin{figure}
\includegraphics[width=0.8\columnwidth]{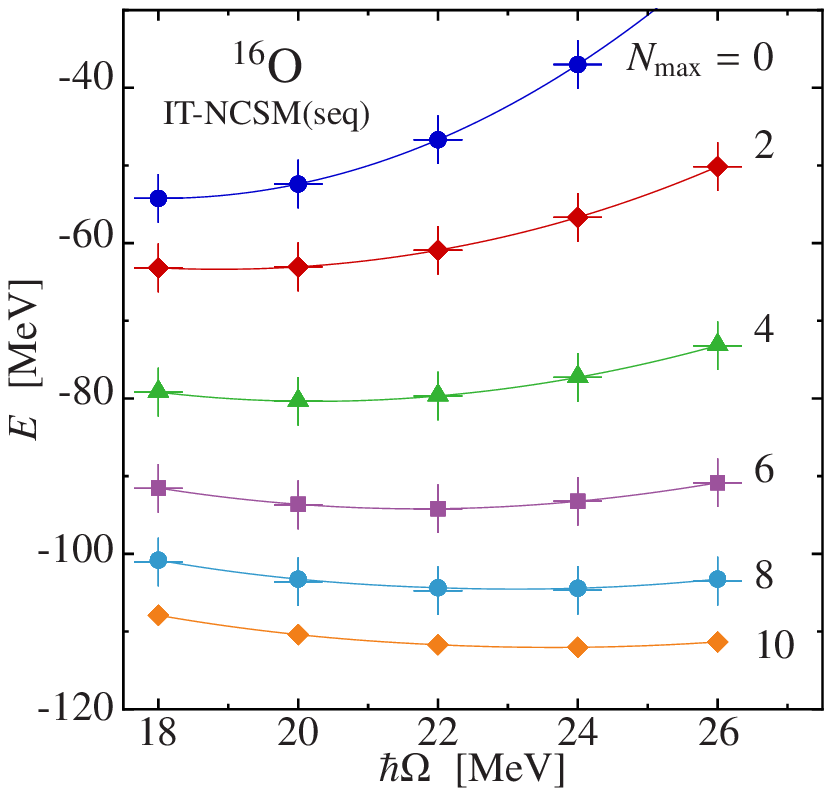}
\caption{(color online) Ground-state energies of \elem{O}{16} obtained for the $\VO_{\UCOM}$ interaction as function of the oscillator frequency $\hbar\Omega$ for different $N_{\max}\hbar\Omega$ model spaces. Shown are the results of IT-NCSM(seq) calculations (solid symbols) in comparison to full NCSM calculations (crosses).}
\label{fig:app_O16_sequential_hwdep}
\end{figure}

The quality of the IT-NCSM(seq) in comparison to the full NCSM is independent of the oscillator frequency $\hbar\Omega$ of the underlying basis. As shown in Fig. \ref {fig:app_O16_sequential_hwdep} both sets of calculations are essentially on top of each other. The maximum deviations are around $300$ keV, with the IT-NCSM tending to higher energies due to its variational character. 

Based on the results of Fig. \ref{fig:app_O16_sequential} we can attempt an extrapolation $N_{\max}\to\infty$. Close inspection of the $N_{\max}$ dependence reveals a non-exponential behavior for large $N_{\max}$ which affects the quality of the extrapolation. This is a property of the $\VO_{\UCOM}$ interaction used here and is not related to the IT-NCSM itself. Similar calculations with other interactions, e.g. the chiral N3LO potential after an Similarity Renormalization Group evolution used in Ref. \cite{NaQu09,NaRo09}, do not have this problem. If we, nevertheless, use the energies for five consecutive values of $N_{\max}$ to perform an exponential extrapolation, the extrapolated energy has a sizable dependence on the chosen window in $N_{\max}$. When using the IT-NCSM(seq) energies in the window $14\leq N_{\max}\leq 22$ we obtain $-133.1$ MeV, for the range $12\leq N_{\max}\leq 20$ we obtain $-132.4$ MeV, and for $10\leq N_{\max}\leq 18$ we get $-130.8$ MeV. In order to arrive at a stable extrapolation for the $\VO_{\UCOM}$ interaction, one would have to go to even larger $N_{\max}$ or use effective model space interactions constructed via a Lee-Suzuki transformation.

\subsection{Center-of-Mass Contamination}

An important advantage of the NCSM is the possibility to exactly separate the intrinsic and the center-of-mass (CM) component of the many-body states. Only in this way a non-spurious description of the translationally-invariant intrinsic state of the nucleus---and all the observables derived from it---is guaranteed. As discussed in Sec.~\ref{sec:itncsm_modelspace}, this property relies on the use of a complete $N_{\max}\hbar\Omega$ model space constructed from a harmonic oscillator single-particle basis. Any other model-space truncation will destroy the formal separability and lead to CM contaminations of the intrinsic states.  

Since the importance truncation reduces the model space to a subset of the full $N_{\max}\hbar\Omega$ space, it might induce a coupling between intrinsic and CM motion and destroy the exact separability. We have to check explicitly that the IT-NCSM eigenstates still exhibit the separation between intrinsic and CM motion.   

A well-known tool to probe the presence and extent of the coupling is an artificial shift of the excitation spectrum of the CM component of the many-body states. Following Gloeckner and Lawson \cite{GlLa74} this can be implemented by adding a harmonic-oscillator Hamiltonian with respect to the CM position $\XOV_{\cm}$ and the CM momentum $\POV_{\cm}$  
\eq{
  \HO_{\cm}
  = \frac{1}{2mA} \POV_{\cm}^2 + \frac{mA\Omega^2}{2} \XOV_{\cm}^2 - \frac{3}{2}\hbar\Omega \;.
}
The modified Hamiltonian
\eq{
  \HO_{\beta}
  = \HO_{\intr} + \beta \HO_{\cm}
}
is then used instead of the intrinsic Hamiltonian \eqref{eq:itncsm_hamiltonian} at all stages of the calculation.

If intrinsic and center-of-mass motion are properly decoupled, then this shift will not affect the intrinsic state whatsoever. The intrinsic ground-state energy, defined via the expectation value $E_{\intr}(\beta) = \matrixe{\Psi_{\beta}}{\HO_{\intr}}{\Psi_{\beta}}$ computed with the eigenstates $\ket{\Psi_{\beta}}$ obtained for $\HO_{\beta}$, has to be completely independent of $\beta$. Any dependence of $E_{\intr}(\beta)$ on $\beta$ signifies an unphysical coupling of the intrinsic state to the CM state of the nucleus.

\begin{figure}
\includegraphics[width=0.75\columnwidth]{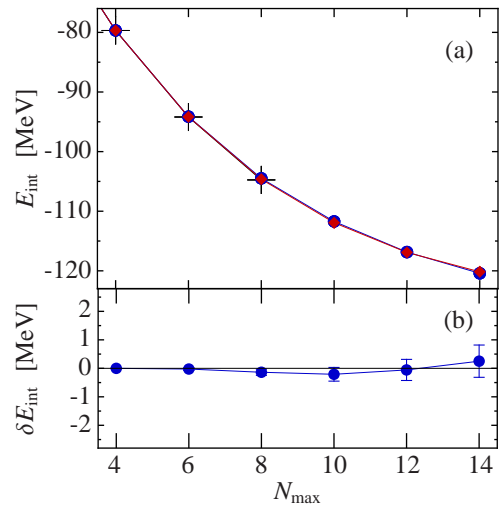}
\caption{(color online) Intrinsic ground-state energy of \elem{O}{16} obtained in the IT-NCSM(seq) at $\hbar\Omega=22$ MeV using the modified Hamiltonian $\HO_{\beta}$. Panel (a) shows the intrinsic energies for $\beta=0$ (\symbolcircle[FGBlue]) and $\beta=10$  (\symboldiamond[FGRed]) in comparison to the full NCSM ($+$). Panel (b) depicts the energy difference $\delta E_{\intr} = E_{\intr}(\beta=10) - E_{\intr}(\beta=0)$. }
\label{fig:app_O16_itncsm_cm}
\end{figure}

As an example for this check, we discuss the \elem{O}{16} ground-state energy obtained in the IT-NCSM(seq) scheme. The IT-NCSM(seq) is set up as described in Sec.~\ref{sec:app_O16}. Fig.~\ref{fig:app_O16_itncsm_cm} depicts the intrinsic energies for a sequence of $N_{\max}$-values obtained for $\beta=0$, i.e. with the intrinsic Hamiltonian used in all previous calculations, and for $\beta=10$. The intrinsic energies of both calculations agree almost perfectly. As shown in Fig.~\ref{fig:app_O16_itncsm_cm}(b) the difference is always below $300$ keV and consistent with $0$ within the uncertainty of the threshold extrapolation. Evidently, the importance truncation does not induce any noticeable coupling between intrinsic and CM degrees of freedom and thus the eigenstates are free of CM contaminations.

The situation is completely different if we start from a model space which is not based on the $N_{\max}\hbar\Omega$ truncation. A well known example is the core-plus-valence-space shell model, where the model space is spanned by Slater determinants generated by all possible occupations of a few valence orbitals. A number of studies show the severity of the problem: As discussed in Ref.~\cite{RaFa90}, e.g., spurious admixtures cause the ground-state energy of \elem{O}{16} to be overestimated by several MeV.  Similar effects are observed when using the importance truncation idea with a no-core model space defined through a truncation of the single-particle basis. These IT-CI calculations, as discussed in Ref. \cite{RoGo08}, exhibit sizable CM contaminations of the intrinsic states which also lead to energy shifts of several MeV for the \elem{O}{16} ground state. A detailed investigation of the CM contaminations in IT-CI and coupled-cluster calculations will be presented elsewhere \cite{PiRo09}.

\section{Applications \& Benchmarks: Non-Magic Nuclei}
\label{sec:app2}

The IT-NCSM is not limited to doubly-magic or closed-shell nuclei. We can apply the same ideas and computational techniques, in particular the IT-NCSM(seq) scheme, without any changes to non-magic or open-shell nuclei. In this section we demonstrate this flexibility and discuss the performance of the IT-NCSM scheme for selected non-magic even-even nuclei from the p-shell in comparison to the full NCSM. A systematic study of p-shell nuclei with different unitarily transformed realistic interactions will be presented in a forthcoming publication.

\subsection{Carbon-12}
\label{sec:app_C12}

As a first step towards open-shell nuclei we consider the ground state of \elem{C}{12} in the IT-NCSM. The computational complexity of this problem is similar to the \elem{O}{16} ground state, because of the incomplete filling of the p-shell. The full NCSM is typically limited to $N_{\max}=8$, whereas the IT-NCSM can be extended to $N_{\max}=22$ and beyond. 

Both schemes for constructing the importance truncated space, the iterative IT-NCSM($i$) and the sequential IT-NCSM(seq) scheme, can be applied without change. For the IT-NCSM($i$) scheme a natural choice for the initial reference state is the ground state obtained from a $0\hbar\Omega$ calculation in the full NCSM instead of the single Slater determinant that spans the $0\hbar\Omega$ space for a magic nucleus. For the IT-NCSM(seq) scheme we start with a full NCSM calculation in a $0\hbar\Omega$ or $2\hbar\Omega$ space in any case, so there is no technical difference between closed- and open-shell nuclei. For brevity, we restrict ourselves to the IT-NCSM(seq) scheme in this section. As in Sec.~\ref{sec:app} we employ a set of calculations with importance thresholds in the range $\kappa_{\min}=3\times10^{-5}$ to $14\times10^{-5}$ for each $N_{\max}$. On this basis we perform a constrained threshold extrapolation as described in Sec.~\ref{sec:itncsm_thresholdextrapol} making use of the second-order perturbative estimate of the energy contribution of excluded configurations. 

\begin{figure}
\includegraphics[width=0.75\columnwidth]{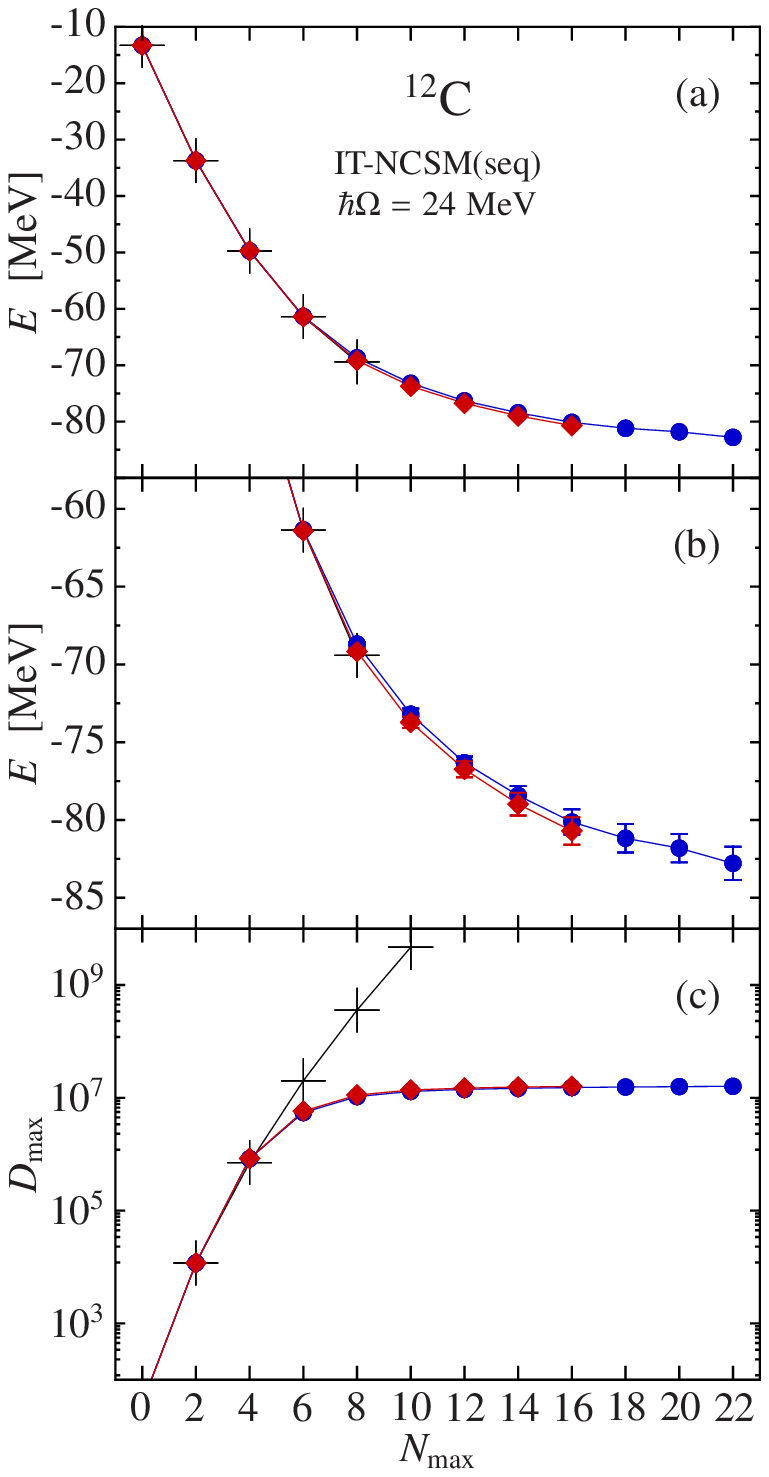}
\caption{(color online) Ground-state energy, (a) and (b), and model-space dimension (c) as function of $N_{\max}$ for \elem{C}{12} obtained within the IT-NCSM(seq) scheme  using the $\VO_{\UCOM}$ interaction for $\hbar\Omega=24$ MeV. Two different values of the parent threshold were used: $C_{\min}=5\times10^{-4}$ (\symbolcircle[FGBlue]) and  $C_{\min}=3\times10^{-4}$ (\symboldiamond[FGRed]). For comparison the results of full NCSM calculations with the same Hamiltonian are included ($+$).}
\label{fig:app_C12_sequential}
\end{figure}

The evolution of the ground-state energy and the model-space dimension with $N_{\max}$ obtained in the IT-NCSM(seq) for the $\VO_{\UCOM}$ interaction is depicted in Fig. \ref{fig:app_C12_sequential}. For the reference threshold we use two different values, $C_{\min}=3\times10^{-4}$ and $5\times10^{-4}$. The sensitivity of the ground-state energy to the reference threshold is slightly larger than for the doubly-magic \elem{O}{16} because of the absence of a single dominant basis state. However, the difference between the two sets of energies remains well below $1$ MeV.

As for \elem{O}{16} the general rate of convergence is rather slow and of non-exponential character for model spaces beyond $N_{\max}\approx14$. To a large extend this can be traced back to the high-momentum behavior of the first-generation $\VO_{\UCOM}$ interaction. A rough extrapolation based on the five data points in the range $14\leq N_{\max} \leq 22$ leads to an estimated ground-state energy of $-84.6(1.5)$ MeV, where the uncertainty is determined by comparing with extrapolations for other sets of five consecutive points. This is almost 8 MeV above the experimental ground state energy of $-92.16$ MeV \cite{AuWa03}. Keeping in mind that the calculated ground-state energy of \elem{O}{16} was at least $5$ MeV below the experimental value, this can be interpreted as evidence for deficiencies in the spin-orbit part of the first generation $\VO_{\UCOM}$ interactions, which in turn could be related to missing three-body interactions.

Interestingly, a similar pattern has been observed for the JISP16 interaction in the full NCSM calculations presented in Ref. \cite{MaVa09}. Although these NCSM calculations were limited to $N_{\max}\leq8$ the softness of the JISP16 interaction allows for quantitative conclusions already in these small spaces. Based on systematic extrapolations the authors conclude that \elem{C}{12} is overbound by approximately 2 MeV and \elem{O}{16} is overbound by $15$ to $18$ MeV. Hence the difference in the binding energies of the two nuclei is of the same order as for the $\VO_{\UCOM}$ interaction although the JISP16 overbinds \elem{O}{16} significantly.

\subsection{Helium-6 and Helium-8}

As a second example we consider the neutron-rich Helium isotopes \elem{He}{6} and \elem{He}{8}. Whereas for \elem{He}{4} one is able to reach large $N_{\max}$ with the full NCSM already, the few additional neutrons in these isotopes significantly reduce the range of the full NCSM, typically to $N_{\max}\leq16$ for \elem{He}{6} and $N_{\max}\leq12$ for \elem{He}{8} \cite{CaNa06}. With the importance truncation we can overcome this limitation easily. 

\begin{figure}
\includegraphics[width=0.75\columnwidth]{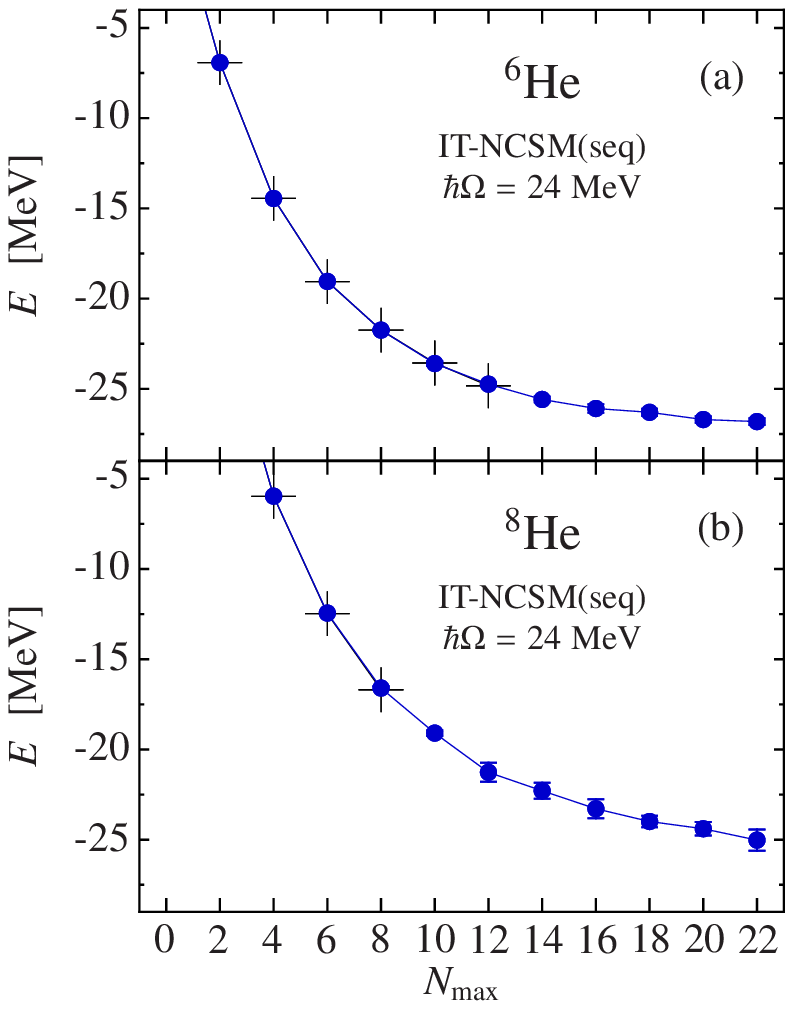}
\caption{(color online) Ground-state energies of \elem{He}{6} (a) and \elem{He}{8} (b) as function of $N_{\max}$ obtained within the IT-NCSM(seq) scheme (\symbolcircle[FGBlue]) using the $\VO_{\UCOM}$ interaction for $\hbar\Omega=24$ MeV. For comparison the results of full NCSM calculations with the same Hamiltonian are included ($+$).}
\label{fig:app_He6He8_sequential}
\end{figure}

The IT-NCSM(seq) results for the ground states of \elem{He}{6} and \elem{He}{8} obtained with $\VO_{\UCOM}$ at $\hbar\Omega=24$ MeV with $C_{\min}=5\times10^{-4}$ are summarized in Fig.~\ref{fig:app_He6He8_sequential}. As before, the IT-NCSM(seq) energies show an excellent agreement with the results of full NCSM calculations where the latter are feasible. For larger $N_{\max}$ the threshold extrapolation shows uncertainties of up to $700$ keV for \elem{He}{8}. If necessary, these uncertainties can be reduced by considering lower $\kappa_{\min}$-values for the threshold extrapolation. The general convergence as a function of $N_{\max}$ is rather slow, particularly for \elem{He}{8}. In addition to the properties of the $\VO_{\UCOM}$ interaction discussed before, the structure of these nuclei affects the convergence rate. Obviously, the description of the neutron halo in an oscillator basis requires high-lying single-particle states and thus large $N_{\max}$. Only through the importance truncation these large model spaces are accessible.

Because of the slow convergence and the relatively large uncertainties of the threshold extrapolation, an extrapolation to $N_{\max}\to\infty$ only provides a rough estimate. Using the results for the five largest spaces we obtain an extrapolated ground-state energy of $-27.4(1.0)$ MeV for \elem{He}{6} and of $-26.5(1.5)$ MeV for \elem{He}{8}. A systematic study including a variation of the oscillator frequency is needed to provide more precise extrapolations. The comparison of these estimates with the experimental binding energies of $-29.27$ MeV and $-31.41$ MeV \cite{AuWa03} for \elem{He}{6} and \elem{He}{8}, respectively, confirms our observations regarding the deficiencies of the first-generation $\VO_{\UCOM}$ interactions. The systematic underbinding of these open-shell systems could be remedied, e.g., by a stronger spin-orbit component of the interaction. Again, the NCSM studies with the JISP16 interactions presented in Ref. \cite{MaVa09} show a similar trend, though the absolute deviations are smaller.

\section{Conclusions \& Outlook}

We have introduced an importance truncation scheme with all its technical aspects as a new tool to facilitate ab initio nuclear structure calculations beyond the domain of conventional CI approaches. Based on an \emph{a priori} importance measure derived from multiconfigurational perturbation theory we identify the important configurations for the description of individual target states such that the dimension of the eigenvalue problem that needs to be solved is dramatically reduced. The effect of excluded configurations can be reliably included by combining a perturbative estimate of their energy contribution with threshold extrapolation techniques. 

In combination with the $N_{\max}\hbar\Omega$ space of the NCSM the importance truncation provides a powerful tool to asses all aspects of nuclear structure in light and medium-heavy nuclei. The importance truncation preserves a crucial property of the NCSM: the decoupling of intrinsic and center-of-mass degrees of freedom which guarantees that the intrinsic observables are free of unphysical center-of-mass contaminations. We have discussed two schemes for setting up the importance-truncated space, the iterative IT-NCSM($i$) and the sequential IT-NCSM(seq) scheme. The latter is most efficient since we have to construct the importance-truncated space only once for each $N_{\max}$. 

Moreover, the IT-NCSM(seq) scheme is conceptually superior, because in the limit of vanishing thresholds $C_{\min}$ and $\kappa_{\min}$ the complete $N_{\max}\hbar\Omega$ space is obtained without any truncation regarding the $npnp$ order at each step of the sequence of $N_{\max}$ values. Hence, the full NCSM results are recovered in the limit $(C_{\min},\kappa_{\min})\to0$ at each $N_{\max}$. Based on this property we use a numerical \emph{a posteriori} threshold-extrapolation to obtain an approximation to the full NCSM with well-defined error bounds. The stability of this extrapolation is greatly enhanced by using information on the contribution of excluded configurations from perturbation theory. Further improvements of these extrapolation techniques, e.g. along the lines discussed in Refs.~\cite{AnCi97} or \cite{ZhNo04}, will be investigated in the future. 

Our series of benchmark calculations confirms the excellent agreement of the IT-NCSM with the full NCSM in all cases where the latter is computationally feasible. The comparison also demonstrates that the IT-NCSM gives access to much larger $N_{\max}\hbar\Omega$ spaces and to heavier nuclei than the full NCSM. The range of the IT-NCSM in both, $N_{\max}$ and $A$ is only limited by the computing time and not by memory. Moreover, the time-consuming steps of the computation can be easily parallelized with minimal communication overhead.

The present calculations also allow for a detailed assessment of the first-generation $\VO_{\UCOM}$ interactions used. Whereas the \elem{He}{4} binding energy is in agreement with experiment by construction, the ground state of \elem{O}{16} is overbound by at least $5$ MeV. This level of agreement is still satisfactory and is not found with most other realistic two-body interactions, be it bare of effective. For the non-magic nuclei discussed here the binding energies are systematically underestimated with the $\VO_{\UCOM}$ interaction, which might hint at deficiencies in the spin-orbit part of the interaction. Furthermore, the IT-NCSM calculations show that the convergence rate of the first-generation $\VO_{\UCOM}$ when going to large spaces is rather slow, which might result from the high-momentum behavior of the interaction. All of these deficiencies will be addressed during the construction of the next generation of UCOM-transformed interactions and the IT-NCSM provides a indispensable tool for assessing these aspects. 

Obviously, the investigation of ground states of closed- and open-shell nuclei is only a first step towards a complete \emph{ab initio} description of nuclear structure. The next crucial step is the extension of the IT-NCSM to excited states. The importance-truncation scheme can be generalized in a straight-forward manner for the simultaneous description of a few target states. In this way it becomes possible to describe, e.g., ground and a few excited states simultaneously and on the same footing. A detailed discussion of the methodical details will be presented in a subsequent paper, together with a variety of applications. 

Since we automatically obtain a representation of the eigenstates in a shell-model basis, all observables of interest can be computed directly. Although we discussed only energies for the purpose of the present benchmark, we have computed a variety of properties, e.g., radii, density distributions, and form factors. We have even used the IT-NCSM eigenstates as input for the calculation of phase-shifts for low-energy nucleon-nucleus scattering reactions in the framework of the NCSM/resonating group method (NCSM/RGM) \cite{QuNa08,NaRo09}.

This demonstrates that the IT-NCSM offers the same possibilities for complete \emph{ab initio} calculations of nuclear structure, spectroscopy, and reactions as the full NCSM. At the same time, the IT-NCSM extends the range of these \emph{ab initio} studies to heavier nuclei and larger model spaces, which is crucial for developing a consistent framework for nuclear structure theory throughout the whole nuclear chart.

\section*{Acknowledgments}

I would like to thank Petr Navr\'atil, Bruce Barrett, Piotr Piecuch, Hans Feldmeier, and Heiko Hergert for numerous fruitful discussions and comments. This work is supported by the Deutsche Forschungsgemeinschaft through contract SFB 634 and by the Helmholtz International Center for FAIR within the framework of the LOEWE program launched by the State of Hesse. I thank the Institute for Nuclear Theory at the University of Washington for its hospitality and the Department of Energy for partial support during the completion of this work.


\end{document}